\documentclass[aps,prb,twocolumn, showpacs,superscriptaddress]{revtex4}
\usepackage{amssymb}
\usepackage{float}

\usepackage{graphicx,amsmath}
\usepackage{epstopdf}
\pdfoutput=1

\newcommand{\bea}{\begin{eqnarray}}
\newcommand{\eea}{\end{eqnarray}}
\newcommand{\beq}{\begin{equation}}
\newcommand{\eeq}{\end{equation}}
\newcommand{\benu}{\begin{enumerate}}
\newcommand{\enu}{\end{enumerate}}
\newcommand{\la}{\langle}
\newcommand{\ra}{\rangle}
\newcommand{\al}{\alpha}
\newcommand{\be}{\beta}
\newcommand{\om}{\omega}

\newcommand{\si}{\sigma}

\newcommand{\ham}{\mathcal{H}}
\newcommand{\han}{\mathcal{N}}

\newcommand{\cda}{c^{\dagger}}
\newcommand{\fda}{f^\dagger}
\newcommand{\bk}{{\bf k}}
\newcommand{\bq}{{\bf q}}

\newcommand{\Tcoh}{T_{\rm coh}}
\newcommand{\bTcoh}{{\bar T}_{\rm coh}}
\newcommand{\TK}{T_{\rm K}}
\newcommand{\nn}{\nonumber \\}
\newcommand{\Tr}{\mbox{Tr}}

\begin{document}
\title{
Kondo lattices with inequivalent local moments: \\
Competitive vs. co-operative Kondo screening
}
\date{\today}

\author{Adel Benlagra}
\affiliation{Institut f\"ur Theoretische Physik, Technische Universit\"at Dresden,
01062 Dresden, Germany}
\author{Lars Fritz}
\affiliation{Institut f\"ur Theoretische Physik, Universit\"at zu K\"oln, Z\"ulpicher
Stra\ss e 77, 50937 K\"oln, Germany}
\author{Matthias Vojta}
\affiliation{Institut f\"ur Theoretische Physik, Technische Universit\"at Dresden,
01062 Dresden, Germany}

\begin{abstract}
While standard heavy fermion metals feature a single spin-$1/2$ local moment per unit cell,
more complicated systems with multiple {\em distinct} local moments have been synthesized as well,
with Ce$_3$Pd$_{20}$(Si,Ge)$_6$ being one example.
Here, we discuss the physics of a Kondo lattice model with two local-moment sublattices,
coupled with different Kondo couplings to conduction electrons. The phase diagram will be
strongly modified from that of the standard Kondo lattice if the characteristic screening
temperatures of the distinct moments are well separated.
Therefore, we investigate the interplay between the two Kondo effects using a local
self-energy approximation via slave bosons. We find that the two Kondo effects can either
compete or co-operate depending on the conduction-band filling. In the regime of
competition, small differences in the two Kondo couplings can lead to huge differences in
the respective Kondo scales, due to non-trivial many-body effects.
We also study the low-temperature properties of the collective heavy Fermi-liquid state
and propose a connection to depleted Kondo lattice systems.
\end{abstract}

\maketitle


\section{Introduction}
\label{sec:introduction}

Inter-metallic compounds containing rare-earth ions have been the subject of intense
experimental and theoretical efforts, owing to their fascinating physical properties.
These include heavy fermion mass enhancement, unconventional magnetism and
superconductivity, non-Fermi liquid behavior as well as the exciting possibility of
quantum criticality beyond the Landau-Ginzburg-Wilson paradigm.\cite{stewart01,hvl,geg08}
Microscopically, most
systems contain one rare-earth ion per unit cell, with a single unpaired $f$ electron or
hole, lending themselves to a theoretical description in terms of a Kondo lattice model
with one spin-1/2 local moment per unit cell.

Recent work \cite{JPSJ78_goto, PRB81_064427_deen, PSSB247_winkler} on the family of
ternary compounds (RE)$_3$Pd$_{20}$X$_6$ (Re=rare-earth, X=Si, Ge), which feature {\em
two} inequivalent Kondo sites per unit cell,\cite{JAC204_gribanov, JAC256_kitagawa,ions_foot}
uncovered a variety of interesting phenomena.
Upon cooling, two magnetic phase transitions occur.\cite{PB259_338_kimura,
JAC306_40_donni, JAC323_509_herr} In Ce$_3$Pd$_{20}$Si$_6$, regarded as one of the
heaviest electron Kondo systems with a low-temperature specific-heat coefficient of
$C/T=8\,\mbox{J}/\mbox{mol Ce K}^2$, the lower transition can be tuned to zero
temperature by application of a magnetic field, resulting in non-Fermi liquid signatures
with a linear temperature dependence of the resistivity.\cite{JPCF51_239_strydom,
JMMM316_90_paschen} Furthermore, substitution studies
\cite{PRB80_prokofiev,PSSB247_winkler} suggest that quantum criticality might also be
achieved by pressure. Together, these results call for theoretical investigations of
Kondo lattices with multiple Kondo ions.

\subsection{General considerations}

Theoretically, little is known about the behavior of Kondo lattices with multiple
distinct local moments per unit cell. Here we start with a few general considerations for
a system with two distinct Kondo ions.

For sufficiently strong Kondo couplings, a paramagnetic Fermi liquid phase with fully
screened moments can be expected. Its Fermi volume is given by ${\cal V}_{\rm FL} = K_d
(n_{\rm tot}\,{\rm mod}\,2)$ with $n_{\rm tot}=2n_c+2n_f = 2n_c+2$,
where $2n_c$ and $2n_f$ denote the average number of $c$ and $f$ electrons per unit cell,
respectively.
Further, $K_d = (2\pi)^d/(2 v_0)$ is a phase space factor, with $v_0$ the unit cell volume
and the factor of two accounting for the spin degeneracy of the bands.
If, in contrast, inter-moment interactions dominate, phases with magnetic long-range
order (LRO, e.g. antiferromagnetic) without Kondo screening appear natural;
alternatively, inter-moment singlet formation of spin-Peierls type can also occur. Given
two spin-1/2 moments per crystallographic unit cell (more generally, an even number),
both phases can be realized without further lattice symmetry breaking, resulting in a
Fermi volume ${\cal V}_{\rm AF} = K_d (2n_c\,{\rm mod}\,2)$ which equals ${\cal V}_{\rm
FL}$.


\begin{figure}[t]
\centering
\includegraphics[width =7.5cm]{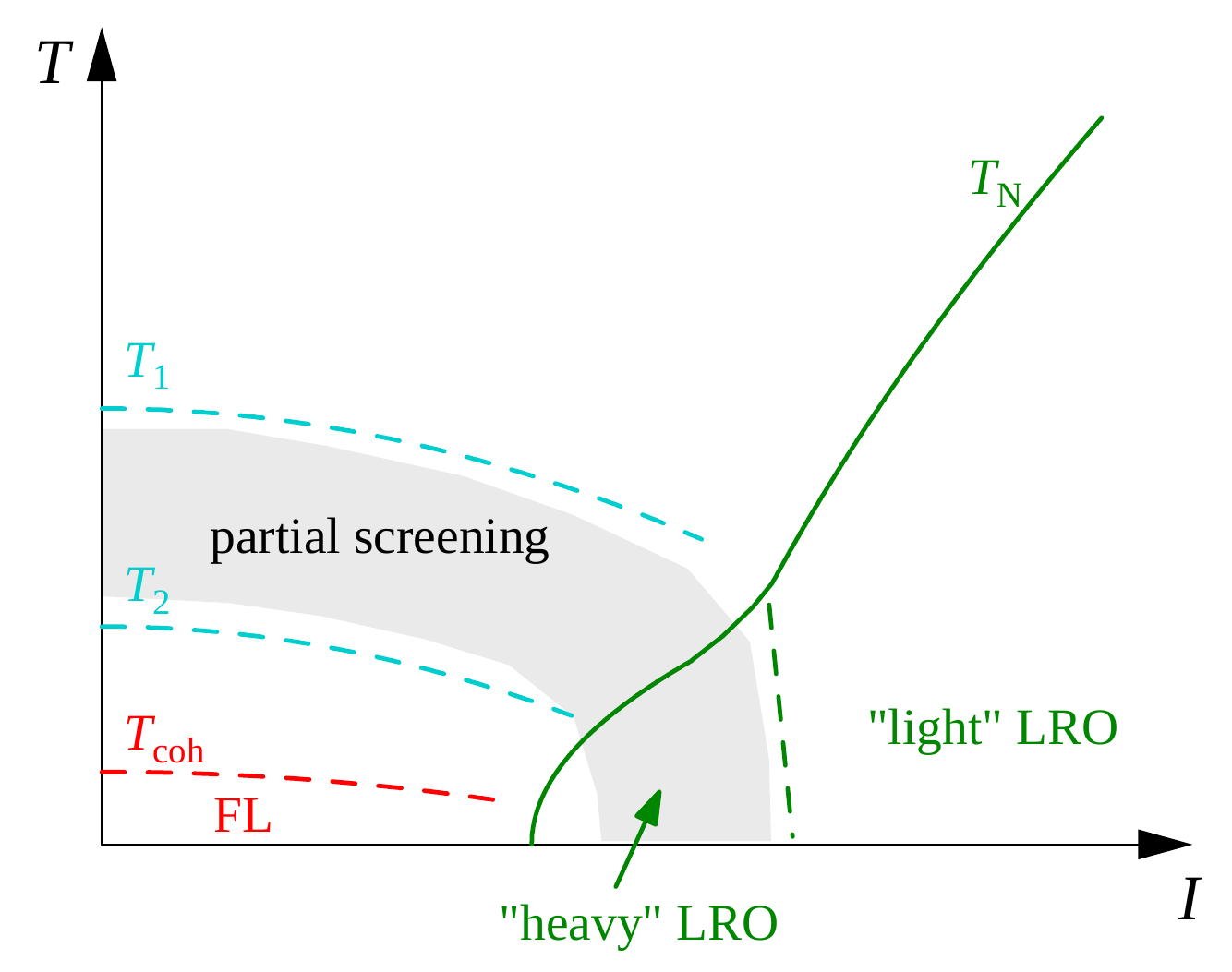}
\caption{ Schematic phase diagram of a Kondo lattice with two inequivalent local moments
per unit cell, as function of temperature $T$ and the magnitude of the inter-moment
exchange $I$ (measured relative to a microscopic Kondo scale, e.g., the largest
single-ion Kondo temperature of the system).
The crossover temperatures $T_{1,2}$ mark the high-temperature onset of screening of the two ions
in their lattice environment, while $\Tcoh$ is the lattice coherence scale below which
the system displays Fermi-liquid behavior. Partial Kondo screening is realized for
$T_2<T<T_1$ (shaded).
Upon increasing $I$, the system enters a long-range-ordered (e.g. antiferromagnetic)
phase where the order is mainly carried by the low-$\TK$ ions (here ``2''), while the
high-$\TK$ ions (``1'') remain almost fully screened -- this is a ``heavy LRO'' phase.
Further increasing $I$ induces a crossover into a more conventional regime where all
moments participate in the order (``light LRO''). It is conceivable that the ``light''
and ``heavy'' LRO regimes are characterized by different ordering patterns, necessitating
further phase transitions inside the LRO phase.
The focus of this paper is on the regime of small $I$.
}
\label{fig:pd}
\end{figure}

Interestingly, a novel regime with {\em partial} Kondo screening may be reached when the
characteristic Kondo scales of the two ions are distinctly different. First, upon
increasing the temperature in the paramagnetic Fermi liquid, the magnetic entropy will be
released in two stages, allowing to define two ``screening'' temperatures $T_1$ and $T_2$
associated with the two ions. Second, upon increasing inter-moment interactions at low
temperature, magnetically ordered phases can occur, where one set of moments is strongly
polarized with little screening, while the second set of moments still forms heavy
quasiparticles with weak polarization. Such a magnetic phase is -- apart from the
possibility of Lifshitz transitions associated with changes in the Fermi-surface topology
-- continuously connected to a local-moment metallic magnet without Kondo
screening,\cite{vojta08} but observables will reflect the co-existence of local-moment
magnetism with heavy fermion behavior. A schematic phase diagram resulting from this
discussion, which extends Doniach's considerations \cite{doniach77} to the case of
multiple distinct Kondo ions, is shown in Fig.~\ref{fig:pd}.

A central question is then under which conditions the two screening scales $T_1$ and $T_2$
will be vastly different. On the one hand, this of course depends on bare model
parameters (i.e. the Kondo couplings and local densities of states), but -- on the other
hand -- there is room for mutual influence and non-trivial interplay of the screening
processes.
The latter can be phrased into the following question:
Does the onset of Kondo screening for one set of moments at some
temperature $T_1$ promote or inhibit the screening of the second set of moments? Naive
arguments can be found for either scenario: If the main effect of the higher-temperature
Kondo screening is to inject additional states near the Fermi level, then the
lower-temperature Kondo effect should be boosted.\cite{proximity} Alternatively, a
variant of the exhaustion argument \cite{exhaus_foot} would imply that, once the
higher-temperature Kondo effect sets in, fewer conduction electrons are left available
for the lower-temperature Kondo effect, which is then inhibited.
In the latter scenario, small differences in the two Kondo couplings can cause largely
different $T_{1,2}$.
Naturally, the lower of the two screening temperatures $T_{1,2}$ sets an upper bound to
the low-temperature coherence scale $\Tcoh$ of the heavy Fermi liquid (experimentally
often defined as the upper boundary of $T^2$ behavior in the resistivity). However, even for standard Kondo
lattices with one local moment per unit cell it has been shown \cite{burdin1,pruschke,grenzebach}
that $\Tcoh$ can be tiny compared to the onset temperature of screening (provided
that inter-site effects are negligible \cite{pines_scaling}), in particular in the so
called ``exhaustion regime`` of small band filling. Clearly, the issue of what determines
the coherence scale $\Tcoh$ in the case of multiple distinct local moments is far from
obvious.

The purpose of this paper is to tackle the simplest set of questions outlined above,
namely that of screening and coherence scales in the heavy Fermi-liquid regime of a
Kondo lattice model with two distinct local moments. These issues can be efficiently
addressed in a local self-energy approximation using slave
bosons.\cite{hewson,pines_scaling} We shall discuss the interplay between the Kondo
crossovers of the two sets of moments. As detailed below, we find that -- depending on
the conduction-band filling $n_c$ -- the Kondo processes may either compete or
co-operate. For both cases, we describe the consequences for the low-temperature behavior
and discuss experimental implications. In particular, we shall argue that small
conduction-band filling tends to open a window of partial Kondo screening as in
Fig.~\ref{fig:pd}. Further, the dominant influence of the conduction-band
filling furnishes a link between the problem studied here and a seemingly different one,
namely a Kondo lattice with random depletion of local moments.\cite{kaul_dep,burdin_dep,ogata_dep}

We note that partial Kondo screening, somewhat similar to the regime advertised in
Fig.~\ref{fig:pd}, has been predicted to occur in certain frustrated Kondo lattices,
where a spontaneous modulation of the Kondo effect is instrumental in releasing geometric
frustration.\cite{lacroix,motome}
\subsection{Outline}
The remainder of this paper is organized as follows.
In Sec. \ref{sec:model} we introduce the microscopic Kondo lattice model with two
distinct Kondo ions, to be studied in this paper. This model is studied in the framework
of the slave-boson mean-field approach. The emergent Kondo scales are derived and
discussed in Sec. \ref{sec:scales}, including the physical picture of competing vs.
co-operating Kondo effects. The low-temperature properties of the Fermi liquid are the
subject of Sec. \ref{sec:FL}. We shall conclude by relating our results to
depleted-Kondo-lattice physics and by discussing directions for future work. Technical
details are relegated to the appendices.


\section{Model and mean-field approximation}
\label{sec:model}

In the body of this paper, we discuss a Kondo lattice model with two inequivalent
local-moment sublattices, with a Hamiltonian of the form
\beq
\mathcal H = -\sum_{\la i, j \ra, \si} t_{ij} \cda_{i \si} c_{j \si} + \sum_{i} J_i \vec S_i \cdot \vec s_i - \mu \sum_{i \si}
\left (\cda_{i \si} c_{i \si} - n_c \right ),
\label{model}
\eeq
in standard notation, with $\vec s_i = \sum_{\si \si'} \cda_{i \si} \vec \tau_{\si \si'}
c_{i \si'} /2$ being the spin density of conduction electrons at site $i$ and $n_c$ the
average number of $c$ electrons per  site. The lattice, with a total number of unit cells
$\han$, is assumed to be bi-partite with sublattices A and B. The Kondo coupling $J_i$
differs for both sublattices and is given by
\bea
J_i = J_1 \delta_{iA} + J_2 \delta_{iB}
\label{J_decomposition}
\eea
where $\delta_{i \al} = 1$ (0) for $i\in\al$ ($i\notin\al$) and $\al = A,B$.
In the following, we will assume $J_1 \geq J_2$.

\begin{figure}[t]
\centering
\includegraphics[width =4.0cm]{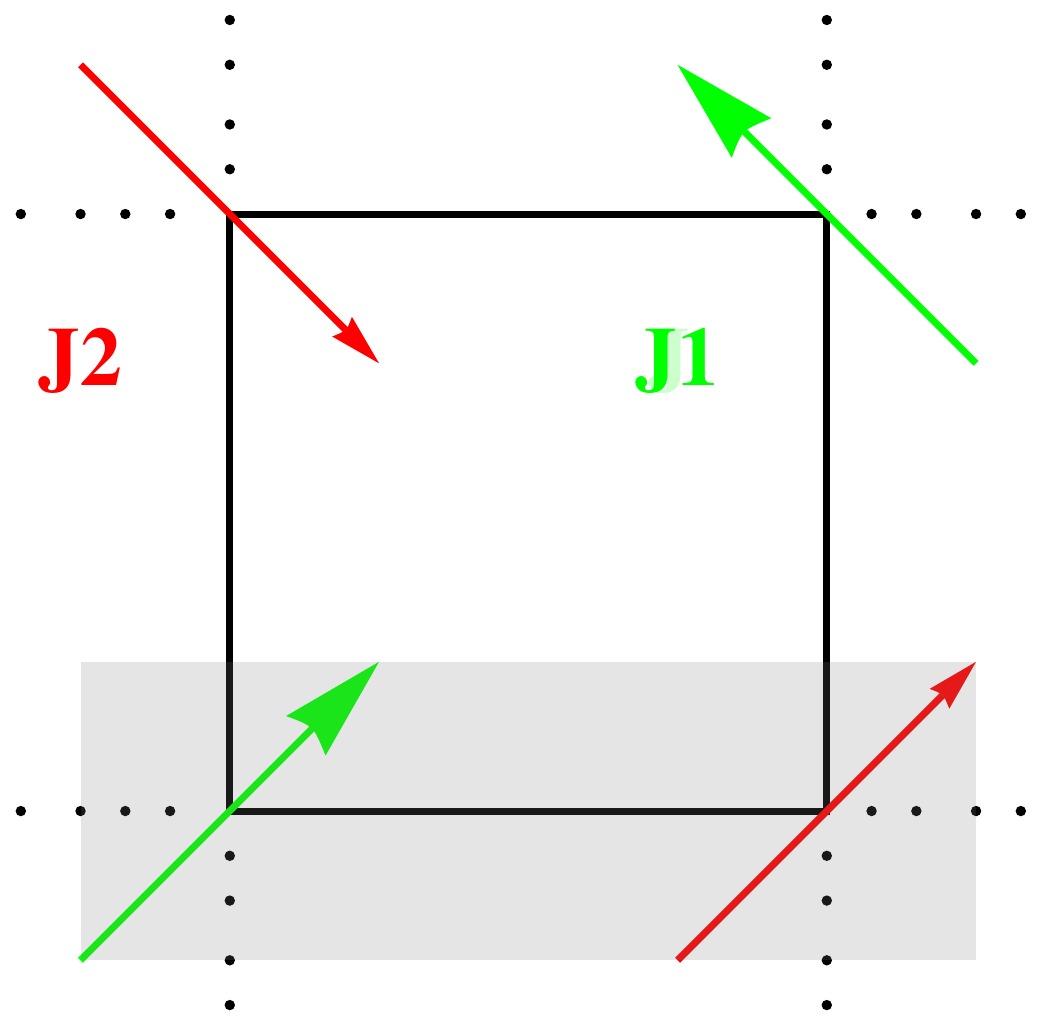}
\caption{
Sketch of a lattice described by the model (\ref{model}). Two types of local
moments are coupled to the conduction electrons of the two sublattices with Kondo
couplings $J_{1,2}$. The shaded area defines the unit cell including two spin-1/2 local
moments. Dots symbolize translation symmetry.
}
\label{lattice}
\end{figure}

The unit cell of the model \eqref{model} contains two $c$ orbitals and two local-moment
(or $f$) orbitals, Fig.~\ref{lattice}. For most of the paper, we shall consider simple
nearest-neighbor hopping only, such that $t_{ij} = t$ on nearest-neighbor bonds and
$t_{ij}=0$ otherwise. Both longer-range hopping and additional kinetic-energy
modulations (compatible with the lattice symmetry) are not expected to qualitatively change
our conclusions; this is explicitly
shown in Appendix \ref{app:square_lattice} for the case of a square lattice with
second-neighbor hopping. For $J_1 = J_2$ our model reduces to the standard Kondo lattice
with conduction electron filling $n_c$, dubbed ``homogeneous'' case in the following.

\subsection{Slave-boson mean-field approximation}
\label{sec:MF}

The model (\ref{model}) will be tackled using a standard slave-boson technique. We
use a fermionic representation for the local spin $\vec S_i$, introducing auxiliary
fermions $f_{i \sigma}$ such that
$$\vec S_i =  \sum_{\si \si'} \fda_{i \si} \vec \tau_{\si \si'} f_{i \si'} /2 ,$$
together with a local constraint $\sum_{\si} \fda_{i \si} f_{i \si} = 1$.
The Kondo interaction in (\ref{model}) now takes the form
\beq
\ham_{K} = -\frac{1}{2} \sum_{i, \si \si'} J_i \fda_{i \si} c_{i \si} \cda_{i \si'} f_{i \si'},
\label{Kondo}
\eeq
where additional terms have can be absorbed into chemical potential and have been thus dropped.
The interaction (\ref{Kondo}) is then decoupled using a field $b_i$ conjugate to $(\sum_{
\si} \sqrt{J_i} \fda_{i \si} c_{i \si})$. The Lagrangian then reads
\bea
\hspace{-0.5cm}\mathcal L &=&T \sum_{i,j, \omega, \si} \left ( -(i \omega + \mu) \delta_{ij} - t_{ij} \right ) \cda_{i \si} c_{j \si}  -  \sum_{i, \si} \lambda_i \fda_{i \si} f_{i \si}   \nn
+&&\hspace{-0.7cm} \sum_{i, \si} \sqrt{J_i} \left ( \fda_{i \si} c_{i \si} b_i + H.c. \right)+ \sum_{ i}  \left (2 b^\dagger_i b_i + \lambda_i \right ) +  2 \han n_c \mu \nn
\label{Final_lagrangian}
\eea
Recall that $\han$ is the number of two-site unit cells of the lattice.

The mean-field theory is obtained from the Lagrangian (\ref{Final_lagrangian}) by
assuming that the fields $b$  and $\lambda$ have uniform and static expectation values on
each sublattice
\beq
\label{approximation}
\langle b_{i \al} \rangle = b_{\al} , ~~\langle \lambda_{i \al} \rangle  = \lambda_{\al} .\nonumber
\eeq
The approximation (\ref{approximation}) is formally exact in the $N\rightarrow
\infty$ limit of a model with the spin symmetry extended to SU($N$).


Performing a Fourier transformation, $c_{\bk \si \al} (f_{\bk \si \al})=
\frac{1}{\sqrt{\han}} \sum_{j\in \al} e^{i\bk.\mathbf{x}_j}c_{j \si} (f_{j\si})$, $\bk$
running over the Brillouin zone of the Bravais lattice,  one obtains the following
mean-field Lagrangian
\bea
\hspace{-0.5cm}\mathcal L & = & - \frac{T}{\han} \sum_{\al, \be} \sum_{\bk, \bq, \omega, \si} \left ( \cda_{\bk \si \al} \fda_{\bk \si \al}\right ) \mathcal{G}^{-1}_{\al \be} (i \omega, \bk) \left ( \begin{array}{c}c_{\bk \si \be}\\ f_{\bk \si \be}\end{array}  \right ) \nn
&+& 2 \han \sum_{\al} b_\al^2 + \han \sum_{\al} \lambda_{\al}+  2 \han n_c \mu,
\label{MFL}
\eea
where the $4\times 4$ inverse Green's function matrix $\mathcal G^{-1}$ is defined by
\beq
\hspace{-0.5cm}\mathcal{G}^{-1}_{\al \be} (i \omega, \bk)  = \left ( \begin{array}{cc}
(i \omega + \mu) \, \, \delta_{\al \be} -  \mathcal E_{\al \be}(\bk) & - \sqrt{J_\al} b_\al \, \, \delta_{\al \be}\\
-\sqrt{J_\al} b_\al \, \, \delta_{\al \be} & (i \omega + \lambda_\al) \, \, \delta_{\al \be}
\end{array} \right ),
\label{Green_matrix}
\eeq
and
\beq
\mathcal E(\bk) = \left ( \begin{array}{cc} \zeta_\bk & \epsilon_\bk^\ast\\\epsilon_\bk &  \xi_\bk
\end{array} \right )
\eeq
is the kinetic-energy matrix whose diagonal (off-diagonal) elements result from intra- (inter-) sublattice hopping.

The mean-field free energy is readily obtained from (\ref{MFL}) as
\bea
F_{MF} &=& - T \,\, \Tr  \log{\left ( - \mathcal{G}^{-1} \right ) }+ 2 \han \sum_{\al} b_\al^2\nn
& +&   \han \sum_{\al} \lambda_{\al} + 2 \han n_c \mu,
\label{FMF}
\eea
where $\Tr$ is a trace over all internal indices $\al, \bk, \omega, \si$.

Self-consistent mean-field equations are obtained by minimizing the free energy
(\ref{FMF}) with respect to $b_\nu$, $\lambda_\nu$ and $\mu$. The result is:
\bea
\label{MFE}
b_\nu&=& \frac{T\sqrt{J_\nu}}{\han} \sum_{\bk, \omega} \mathcal G_{\nu \nu}^{fc}(i \omega, \bk),\nn
1&=&\frac{2 T}{\han} \sum_{\bk, \omega} \mathcal G_{\nu \nu}^{ff}(i \omega, \bk) e^{i\om 0^+},\\
n_c&=&\frac{n_{c1}+n_{c2}}{2},\nonumber
\eea
where the average occupation number of the $c$ band on each sublattice is given by
\beq
n_{c \nu} = \frac{2 T}{\han}\sum_{\bk, i \omega} \mathcal G_{\nu \nu}^{cc}(i \omega, \bk) e^{i\om 0^+}.
\eeq
Here $\mathcal G^{cc}$,  $\mathcal G^{ff}$ and $\mathcal G^{fc}$ are the $2\times2$
sub-matrices of $\mathcal G$, corresponding to the $c$, $f$ and mixed sectors, respectively.
The factors of two in the occupation numbers arise from summation over spin.

In the slave-boson mean-field approximation, the original problem has thus been
mapped onto a model of \emph{four} bands with inequivalent hybridization $\sqrt{J_\al}
b_\al$ between the $c_\al$ and $f_\al$ bands. The resulting renormalized
bands are defined by their dispersions $E_{\bk m}$, $m=1,2,3,4$,
solutions of the eigenvalue problem
\beq
\mbox{Det}\,\mathcal{G}^{-1}(E_\bk,\bk)=0.
\label{eq:eigen}
\eeq

While it is not possible, in general, to obtain a closed form for the solution of
\eqref{eq:eigen}, simplifications arise upon considering inter-sublattice (e.g.
nearest-neighbor) hopping only, such that $\zeta_\bk=\xi_\bk=0$. Further, if sublattice
modulations of the hopping are absent, $\epsilon_\bk$ is real, and the $c$-electron kinetic
energy is given by $\pm\epsilon_\bk$. In this case, one can invert the relation $E\equiv
E_m(\epsilon_\bk)$ and write
\beq
\epsilon(E)=\sqrt{\prod_{\nu=1}^2\frac{J_\nu b_\nu^2
-(\lambda_\nu +E)(\mu+E)}{\lambda_\nu +E}}. \label{eq:inverse_eigen}
\eeq
In particular, at the Fermi level, $E=0$, we have
\beq \mu_L\equiv \epsilon(0) = \sqrt{\prod_{\nu=1}^2
\left (\frac{J_\nu b_\nu^2 }{\lambda_\nu }-\mu \right )}.
\eeq
$\mu_L$ may be interpreted as a $c$-band chemical potential accounting for the
enlargement of the Fermi volume due to the Kondo effect, see Eq.~\eqref{mulint}.
In our case, $b_1, b_2\neq 0$ at $T=0$, and the resulting Fermi volume
${\cal V}_{\rm FL}$ encompasses $(2n_c+2)\,{\rm mod}\,2$ electrons.

In the mean-field equations, the restriction to inter-sublattice hopping only allows one
to convert the momentum summations into energy integrals $$\frac{1}{\mathcal N}\sum_\bk
... \rightarrow \int d\om \rho_0(\om) ...  \,\,\, ,$$ where $\rho_0(\om)$ is the bare
density of states (DOS) of the conduction electrons,
\beq
\rho_0(\om)\equiv1/\mathcal N \sum_\bk  \left [\delta(\om-\epsilon_\bk)+ \delta(\om+\epsilon_\bk) \right ].
\eeq
The DOS is symmetric, $\rho_0(\om) = \rho_0(-\om)$, due to $\zeta_\bk=\xi_\bk=0$.
Note that without sublattice modulations of the bare kinetic energy, $\rho_0$ is
identical to the bare local $c$-electron DOS on both sublattices A and B.

In the following, the equations assume nearest-neighbor hopping only. To ease the numerical implementation, the following calculations have been performed with
either a semi-elliptic DOS, corresponding to a Bethe lattice with nearest-neighbor
hopping, or a constant DOS, with no qualitative differences regarding the interplay
between the two Kondo effects. In addition, we show some results for a square lattice,
with and without next-neighbor hopping hopping, in Appendix \ref{app:square_lattice}.

Finally, we note that non-local two-particle effects are not included in the present
slave-boson approximation. In particular, there are no correlations between different
local-moment spins, i.e., effects of RKKY interaction are neglected. At the mean-field
level, this can be (partially) repaired by explicitly adding an inter-moment interaction
which is then decoupled in the particle-particle or particle-hole channel. Such a
procedure can, e.g., account for a spin-liquid phase competing with Kondo screening, but
this is left for a future study.

\section{Kondo scales}
\label{sec:scales}

Let us start by recalling a few features of the slave-boson approximation applied to Kondo
physics. The onset of Kondo screening is in general signaled by an (artificial) phase
transition: For $T>\TK$, the mean field $b$ vanishes while it is non-zero for $T<\TK$.
For a single Kondo impurity coupled by an exchange coupling $J$ to a bath with a DOS
$\rho_0(\om)$, the mean-field Kondo temperature $\TK(J)$ is the solution of
\beq
\frac{2}{J} = \int_{-\infty}^{+\infty}\!\!d\omega\,\,\frac{\rho_0(\omega+\mu)}{\omega}\tanh{\left
( \frac{\omega}{2 \TK}\right )}.
\label{TKJ}
\eeq
For a constant DOS and in the
weak-coupling limit, $\TK(J)$ evaluates to
\beq
\TK(J) \sim \sqrt{D^2-\mu^2}
\exp\left[-1/(J\rho_0)\right].
\label{exp_Tk}
\eeq
%
Compared to the exact single-ion Kondo temperature, $\TK^{(1)}$, this mean-field $\TK$
has the correct exponential dependence on $J$ and $\rho_0$, and the prefactor matches the
one-loop result for $\TK^{(1)}$. The artificial phase transition at $\TK(J)$ will turn
into the physically correct crossover upon including gauge fluctuations in the
slave-particle description.

Remarkably, the slave-boson approximation applied to a standard Kondo lattice, with bare $c$
electron DOS $\rho_0(\om)$ and Kondo coupling $J$, leads to an onset of screening at exactly
the same $\TK(J)$, simply because $c$ electrons and local moments are
decoupled for $T>\TK(J)$. However, coherent Fermi-liquid behavior may set in at a much
lower temperature $\Tcoh$ (even at the mean-field level), due to protracted screening.\cite{burdin1}

With two inequivalent ions, there will be two onset temperatures for screening, $T_1$ and
$T_2$, i.e., upon increasing the temperature the mean field parameter $b_\nu$ will vanish
at the temperature $T_\nu$. From $J_1>J_2$ we expect $T_1>T_2$. The regime $T>T_1$
consists of conduction electron decoupled from all local moments -- this is a
perturbative high-temperature regime. When $T_2<T<T_1$, the local moments on sublattice A
are hybridized with the $c$ electrons, while those on B are free -- this is the regime of
partial screening advertised in the introduction. Finally, below $T_2$, all local moments
are hybridized. As above, the coherence scale $\Tcoh$, to be discussed in
Sec.~\ref{sec:FL}, may be significantly below $T_2$.

\subsection{Two-stage screening}
The Kondo scales $T_1, T_2$ are readily obtained from the mean-field equations
(\ref{MFE}) in the limits $b_1\rightarrow0$, $b_2 \rightarrow 0$, respectively.
It is then easy to see that
\beq
T_1 = \TK(J_1)
\eeq
with $\TK(J)$ in Eq.~\eqref{TKJ}, because all moments are decoupled for $T>T_1$.
In contrast, $T_2$ has to be determined from
\beq
\frac{2}{J_2} = \int_{-\infty}^{+\infty}d\omega\frac{\rho_{c2}(\omega,
T_2)}{\omega}\tanh{\left ( \frac{\omega}{2 T_2}\right )},
\label{Kondo_scales}
\eeq
where $\rho_{c2}(\om,T)$ is the {\em local} DOS of $c$ electrons on sublattice B. While
$\rho_{c2}(\om)$ is simply given by $\rho_0(\om+\mu)$ for $T>T_1$, the onset of Kondo
screening on the A sublattice renormalizes $\rho_{c2}$ in a temperature-dependent
fashion, which is captured by the following equation:
\begin{equation}
\rho_{c2}(\omega, T)=\alpha(\omega, T)\rho_0 \big(|\omega+\mu|\alpha(\omega, T) \big).
\end{equation}
Here, $\alpha(\omega, T)$ is a ``boost factor'', whose general form is given in appendix
\ref{app:spectral_densities}. For $T>T_2$ it reduces to
\beq
\alpha(\omega, T)=\sqrt{\frac{-J_1
b_1(T)^2}{(\lambda_1(T)+\omega)(\mu(T)+\omega)}+1}.
\label{boost_factor}
\eeq


Due to the non-trivial temperature and frequency dependence of $\rho_{c2}$, the screening
scale $T_2$ will not have the usual form (\ref{exp_Tk}) for the Kondo scale as a function
of the Kondo coupling $J_2$. Still, we expect that an increase of density of states at the
Fermi level $\omega=0$ to enhance this scale. We have, omitting the explicit
temperature dependence of the mean-field parameters
\beq
\rho_{c2}(0)=\sqrt{\frac{-J_1b_1^2}{\mu \lambda_1}+1}\,\,\rho_0 \left
(|\mu| \sqrt{\frac{-J_1b_1^2}{\lambda_1 \mu}+1}\right ).
\label{rhoc2Fermi}
\eeq
If $\rho_{c2}(0)/\rho_0(\mu)>1$, the density of states at the Fermi
level is enhanced by the first Kondo screening and we expect, accordingly, a boost of the
B-sublattice Kondo screening, i.e., $T_2(J_2) > \TK(J_2)$. Alternatively, if
$\rho_{c2}(0)/\rho_0(\mu)<1$, the B-sublattice Kondo screening will be inhibited by the onset
of the A-sublattice one.
In general, the effect of $\alpha(\om,T)$ is two-fold: as a factor affecting the argument
of $\rho_0(\om)$ (which may induce a particle--hole asymmetry) and as an overall
coefficient. For a constant bare density of states, used for most of the numerical
calculations, the first effect is absent. Otherwise, it will induce quantitative changes
depending on the shape of $\rho_0$ -- this happens, e.g., in the square-lattice case in
Appendix \ref{app:square_lattice}.

\begin{figure}[t]
\centering
\begin{tabular}{cc}
(a)&\includegraphics[width = 5.6cm]{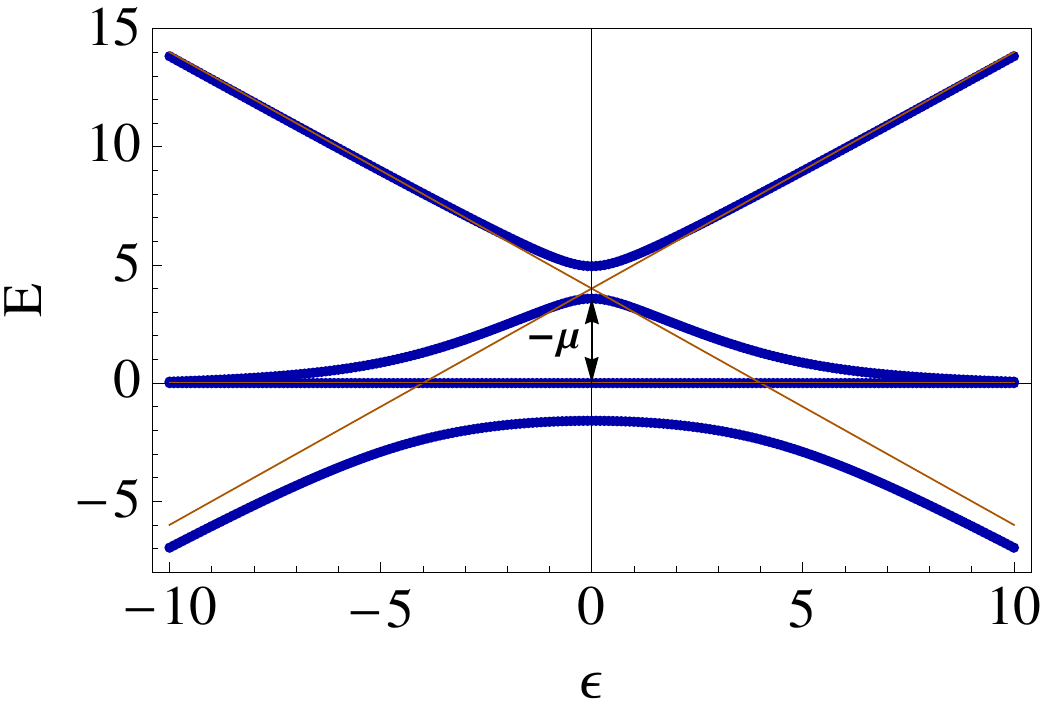}\\
(b)&\includegraphics[width = 5.6cm]{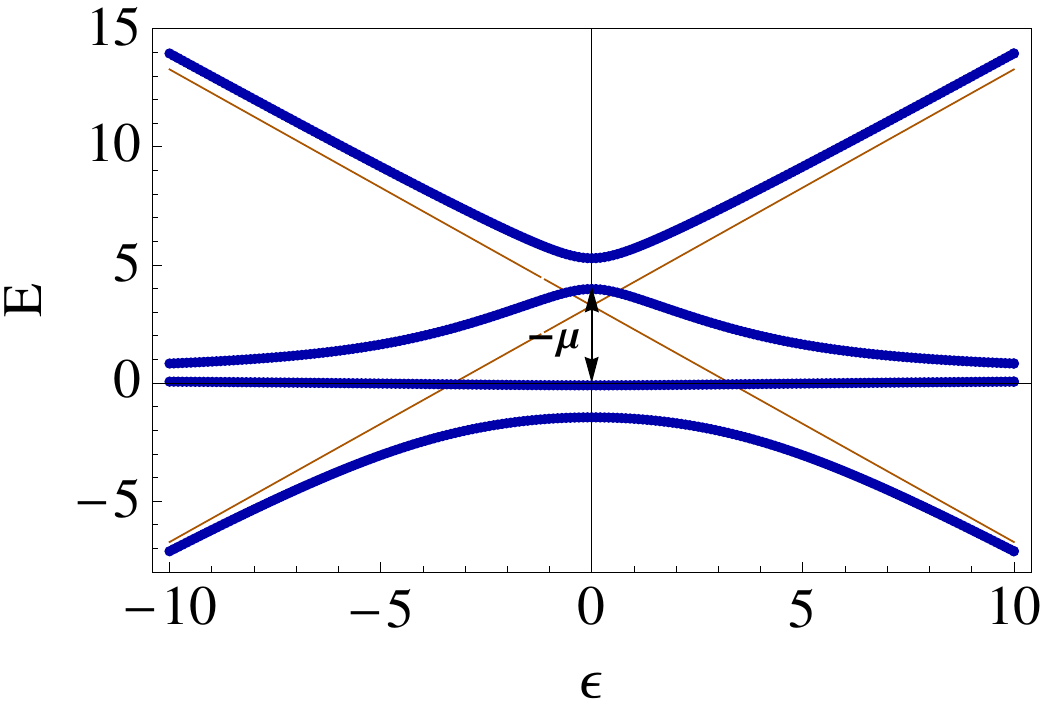}
\end{tabular}
\caption{
Renormalized dispersions $E(\epsilon)$ (thick line) and bare dispersions (thin line) as a
function of $\epsilon$ obtained from equation (\ref{eq:eigen}) for (a) $T_2<T<T_1$ and
(b) $T<T_2$. The bare $c$-electron DOS is constant, $\rho_0=1/(2D)$, with a
half-bandwidth $D=10$.
}
\label{disp_D10_N06}
\end{figure}

Before discussing numerical results for the screening scales $T_{1,2}$, it is
enlightening to look at the renormalized dispersions $E(\epsilon)$ as obtained from
Eq.~\eqref{eq:eigen} by solving the set of mean-field equations. Fig.~\ref{disp_D10_N06}
displays the generic resulting bands for $T_2 < T < T_1$ and $T<T_2<T_1$.
For temperatures $T_2<T<T_1$ (Fig.~ \ref{disp_D10_N06}a), the local moments of the B
sublattice are unscreened; in the mean-field approach, this is signaled by a flat band
being located at the Fermi level $\omega=0$. Hence, three renormalized bands form, one of
which is narrow and primarily of $f$ character. If only nearest-neighbor hopping is
considered, the symmetry of the DOS imposes that at half-filling, $n_c=1$ at $\mu=0$,
this narrow band is completely flat and located at $\omega=0$. Away from half-filling, the
width of this narrow band is then linear in $|\mu|$. With an asymmetric DOS, the width of
the narrow band still reaches its minimum at half-filling (but does not vanish). The
dependence of this width on the chemical potential $\mu$ indicates a strong effect of the
conduction-band filling $n_c$ on the B-sublattice Kondo screening.
Lowering the temperature further below $T_2$ (Fig.~ \ref{disp_D10_N06}b), a
hybridization gap opens near the Fermi level, and the second $f$-level transforms into a
weakly dispersive heavy band crossing the Fermi level.
\begin{figure}[t]
\centering
\begin{tabular}{cc}
(a)&\includegraphics[width = 7.cm]{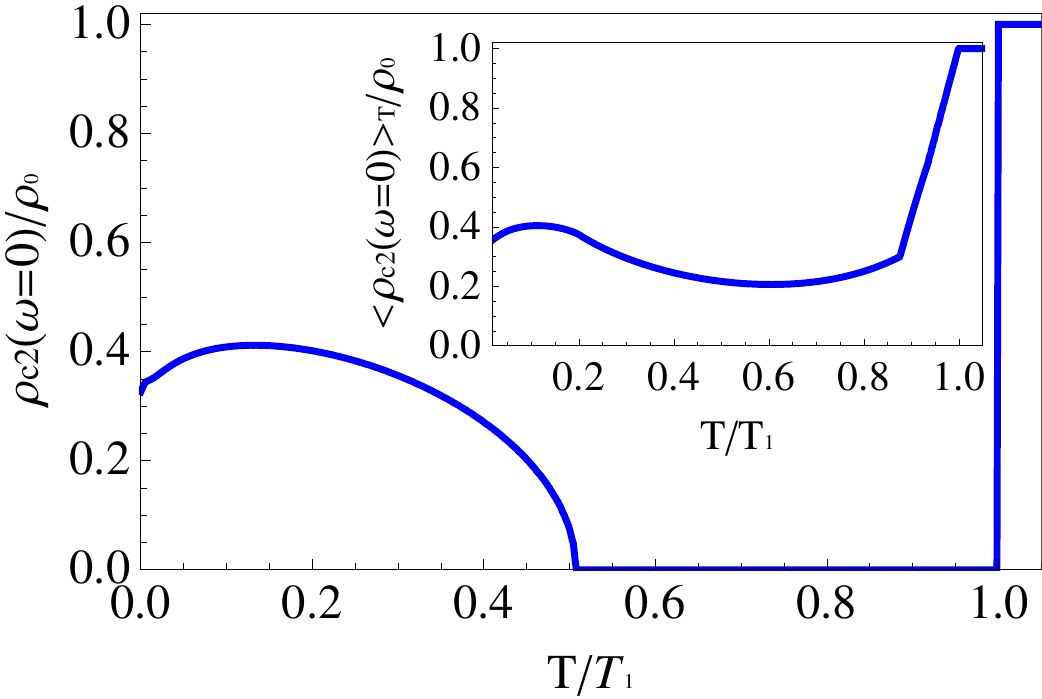}\\
(b)& \includegraphics[width = 7.cm]{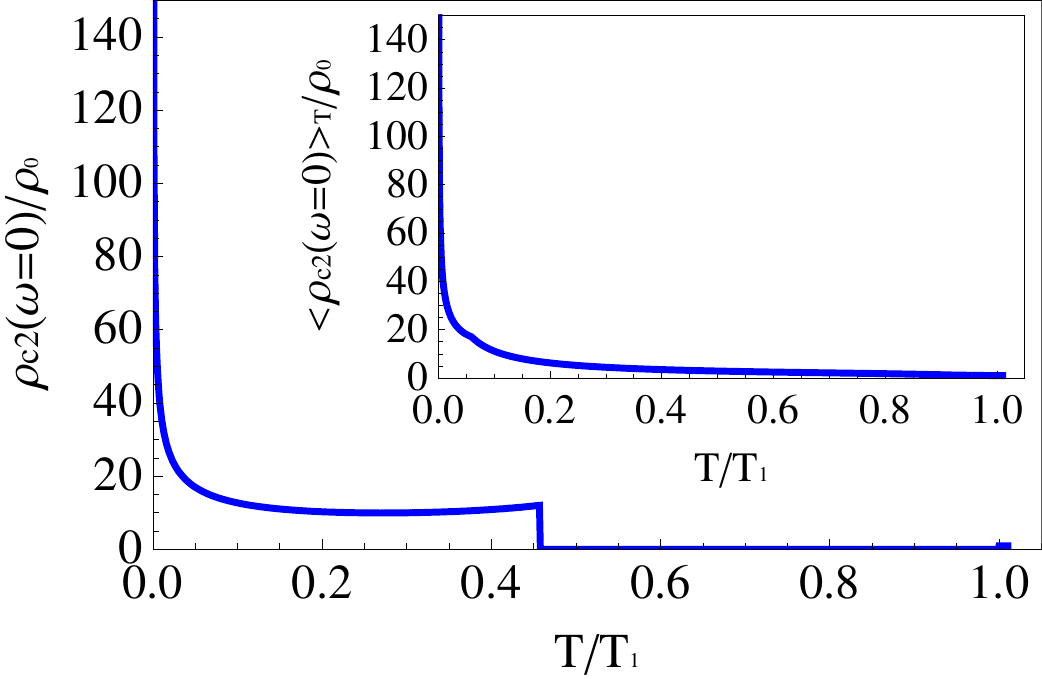}
\end{tabular}
\caption{
Temperature evolution of the normalized density of states at the Fermi level
$\rho_{c2}(0)/\rho_0(\mu)$ (see Eq. (\ref{rhoc2Fermi})) of $c$ electrons in the second
sublattice for $J_2=0$ at an electronic filling number (a) $n_c=0.2$, (b) $n_c=0.9$. The
temperature scale is normalized by $T_1$, the first Kondo coupling is $J_1=8$ and the
half-bandwidth is set to $D=10$. The bare density of states $\rho_0$ is taken constant in
its support. Inset: averaged normalized density of states $\langle\rho_{c2}(0)\rangle_T/\rho_0(\mu)$
in a window of width $\sim T$ around the Fermi level.
}
\label{rho2Fermi_D10}
\end{figure}

\subsection{Competitive vs. co-operative Kondo screening}
\label{sec:fill}

We are now in the position to discuss whether the Kondo effects on the two sublattices
compete or co-operate. To quantify this effect, we shall compare the screening-onset
temperature $T_2$ of the B-sublattice moments in our model with Kondo couplings $J_1>J_2$
to the screening-onset temperature of a homogeneous Kondo lattice with the same bare DOS where all
Kondo couplings are set to $J_2$ -- the latter temperature is simply given by $T_K(J_2)$
according to Eq.~\eqref{TKJ}.

As $T_2$ is determined by Eq.~\eqref{Kondo_scales}, we first focus on the relevant
B-sublattice local DOS $\rho_{c2}(\om,T)$. Fig.~\ref{rho2Fermi_D10} shows the temperature
evolution of the normalized DOS at the Fermi level $\rho_{c2}(\om=0)/\rho_0(\mu)$ for different values of
the electronic filling $n_c=0.2$ and 0.9. Note that we have formally set $J_2=0$, such that all temperatures
obey $T>T_2$.
For both cases, $\rho_{c2}(0)$ vanishes in an intermediate-temperature regime below
$T_1$, as none of the quasiparticle bands crosses the Fermi level. The system is still
metallic: the averaged density of states in a window of width $\sim T$ around the Fermi
level is finite, see the insets of Fig.~\ref{rho2Fermi_D10}. At very low temperatures, we
recover a finite density of states $\rho_{c2}(0)$, as we expect for a metallic state,
except at quarter filling, $n_c=0.5$, where $\rho_{c2}(0)=0$ at all temperatures below $T_1$ and a band insulator is obtained. This can be understood using a counting argument: at
quarter-filling, and for $J_2=0$, the large Fermi surface encompasses an even number $2
n_c+1=2$ of particles and a band gap at the Fermi level is expected.

The crucial difference between Figs.~\ref{rho2Fermi_D10}a and \ref{rho2Fermi_D10}b is
that for $n_c=0.2$, and generically below quarter-filling, $\rho_{c2}(0)$ is smaller than
$\rho_0(\mu)$ at low temperature. Accordingly, we expect the B-sublattice Kondo effect to be
reduced as compared to the homogeneous case. In contrast, for $n_c=0.9$, and generically
above quarter-filling, $\rho_{c2}(0)$ is strongly enhanced. This enhancement reaches its maximum at half-filling where the width of the narrow band reaches its minimal value as discussed earlier. Accordingly, we expect a
sensitive boost of the second Kondo effect when $J_2 \ll J_1$. At quarter-filling, because the
density of states at the Fermi level vanishes at all temperatures below $T_1$,  a
second Kondo screening occurs only above a finite critical value $J_{2c}$.

\begin{figure}[t]
\centering
\begin{tabular}{cc}
(a)&\includegraphics[width = 7.5 cm, height=4.3cm]{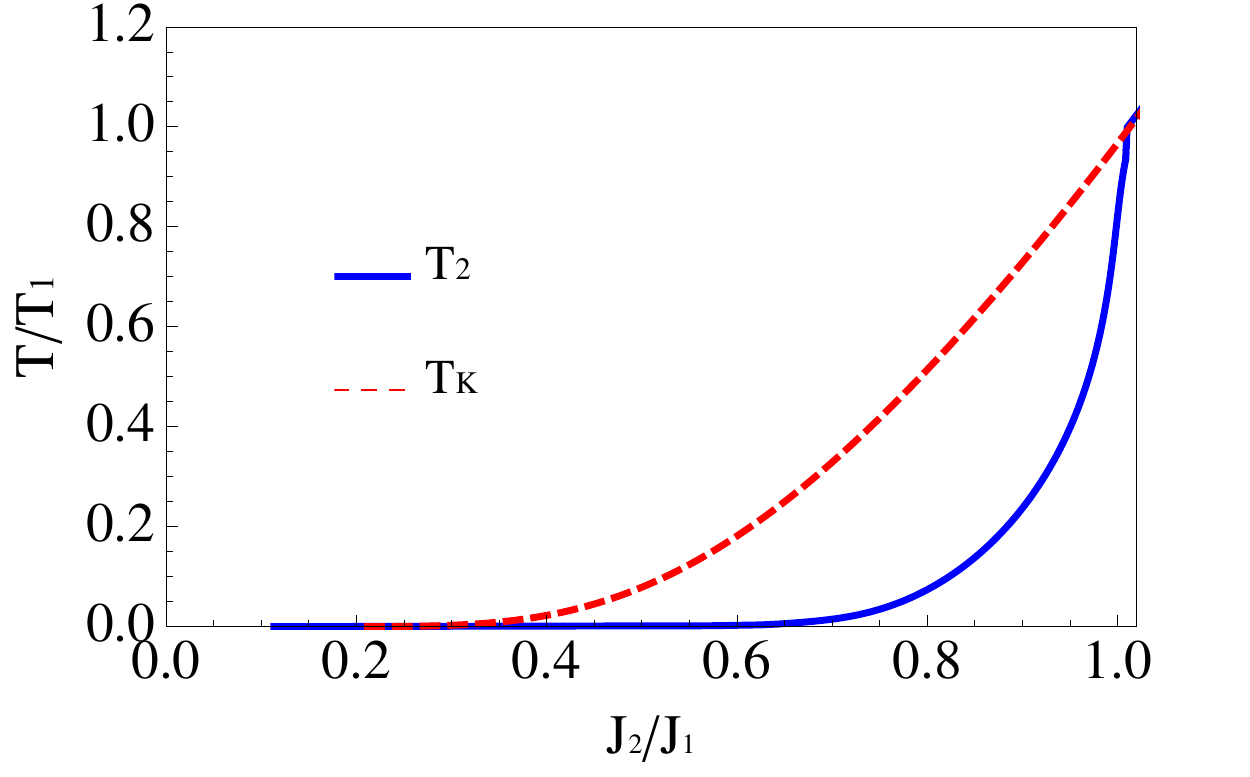}\\
(b)&\includegraphics[width = 7.5 cm, height=4.3cm]{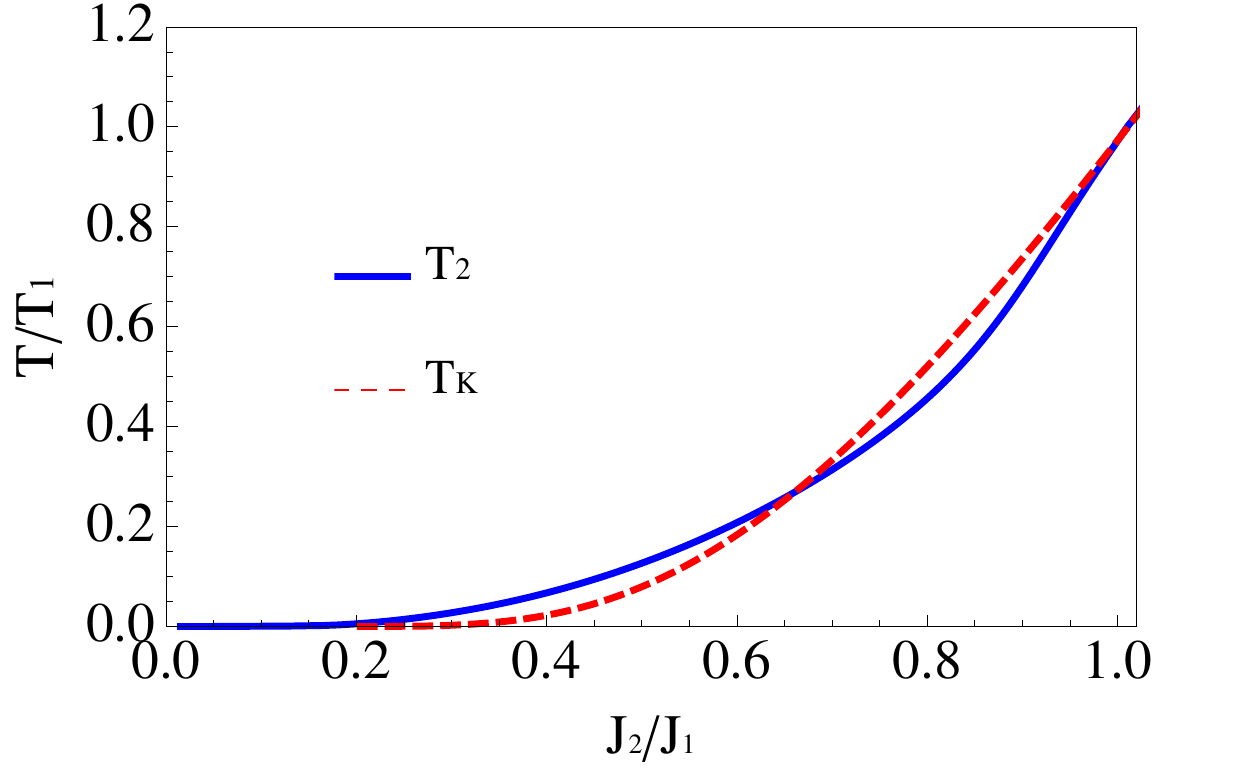} \\
(c)&\includegraphics[width =7.5 cm, height=4.3cm]{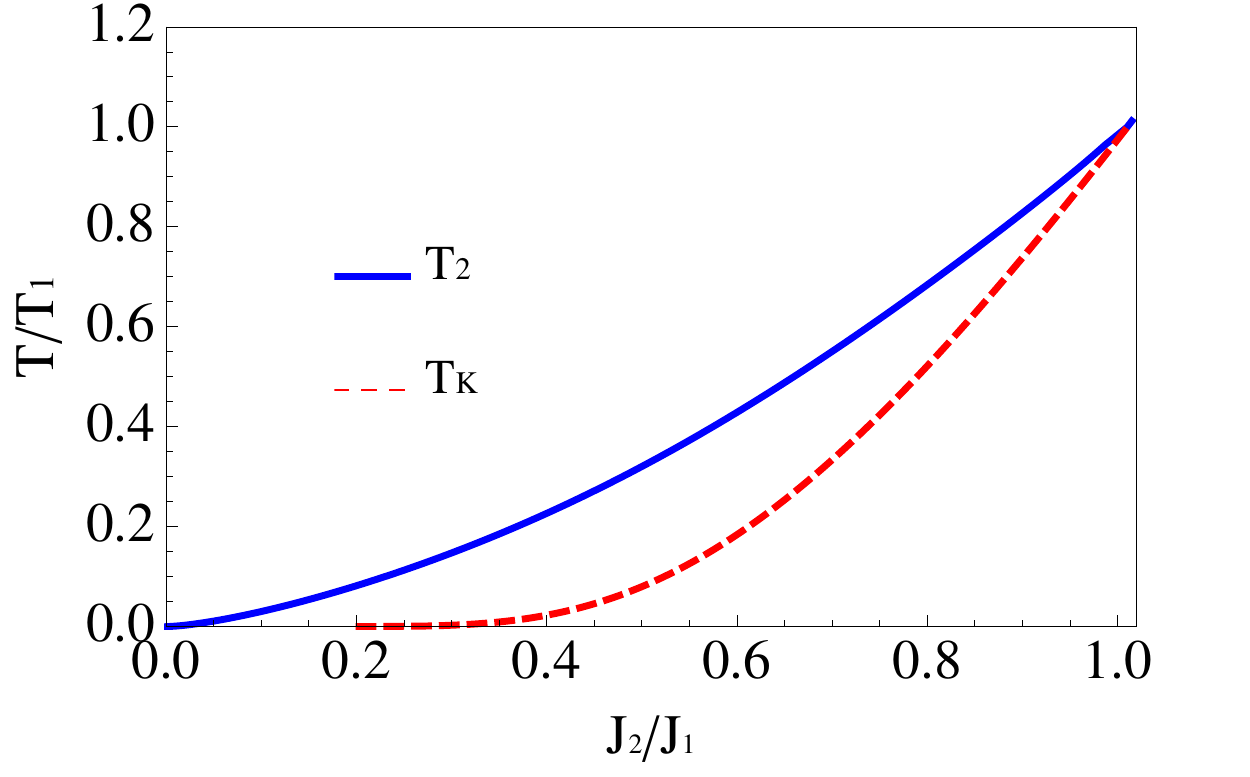}
\end{tabular}
\caption{
Evolution of the temperature scales $T_2(J_2), \TK(J_2)$, where $\TK(J_2)$ is the Kondo
scale obtained in the homogeneous Kondo lattice with a Kondo coupling $J_2$, with the
strength of the second coupling constant $J_2 \leq J_1$ for an electronic filling number:
a) $n_c=0.2$, (b) $n_c=0.6$ and (c) $n_c=0.9$. The temperature scale is normalized by
$T_1$, and the one of $J_2$ by $J_1$. The bare density of states $\rho_0$ used is
constant in its support.
}
\label{Tk_J_D10}
\end{figure}
The B-sublattice screening scale $T_2$, as obtained from Eq.~\eqref{Kondo_scales}, is
displayed in Fig.~ \ref{Tk_J_D10} for different values of the conduction-band filling $n_c$.
At all fillings, the strength of the second Kondo scale is consistent with the
expectations drawn from Fig.~\ref{rho2Fermi_D10} for $J_2 \ll J_1$:
a reduction (enhancement) of the B-sublattice $c$-electron local DOS results in a
suppression (promotion) of the B-sublattice Kondo effect.

When $ n_c \ll 1/2 $, the second Kondo effect is inhibited for all values of $J_2 < J_1$
(Fig.~\ref{Tk_J_D10}a), whereas it is promoted for $1/2 \ll n_c < 1$ for all values of
$J_2 < J_1$ (Fig.~\ref{Tk_J_D10}c). It is worth emphasizing that the suppression of $T_2$
for small $n_c$ is a drastic effect: Even small differences between $J_1$ and $J_2$ can
induce largely different $T_1$ and $T_2$, Fig.~\ref{Tk_J_D10}a. Close to quarter-filling,
a crossover regime is found for which the B-sublattice Kondo effect is promoted for $J_2
\ll J_1$ and inhibited for intermediate values of $J_2$ up to $J_1$
(Fig.~\ref{Tk_J_D10}b). For these intermediate values of $J_2$, it is not straightforward
to deduce the effect of the A-sublattice Kondo effect on the B-sublattice one because the
density of states at the Fermi level $\rho_{c2}(0)$ is vanishing in an intermediate
energy regime. One has then to consider the full integral in \eqref{Kondo_scales} instead
of focusing on its usually dominant $\omega=0$ contribution. We found this transition
regime to depend on the shape of the bare DOS.

\subsection{Kondo feedback}

\begin{figure}[tb]
\vspace{-0.cm}
\centering
 \includegraphics[width = 8.5cm]{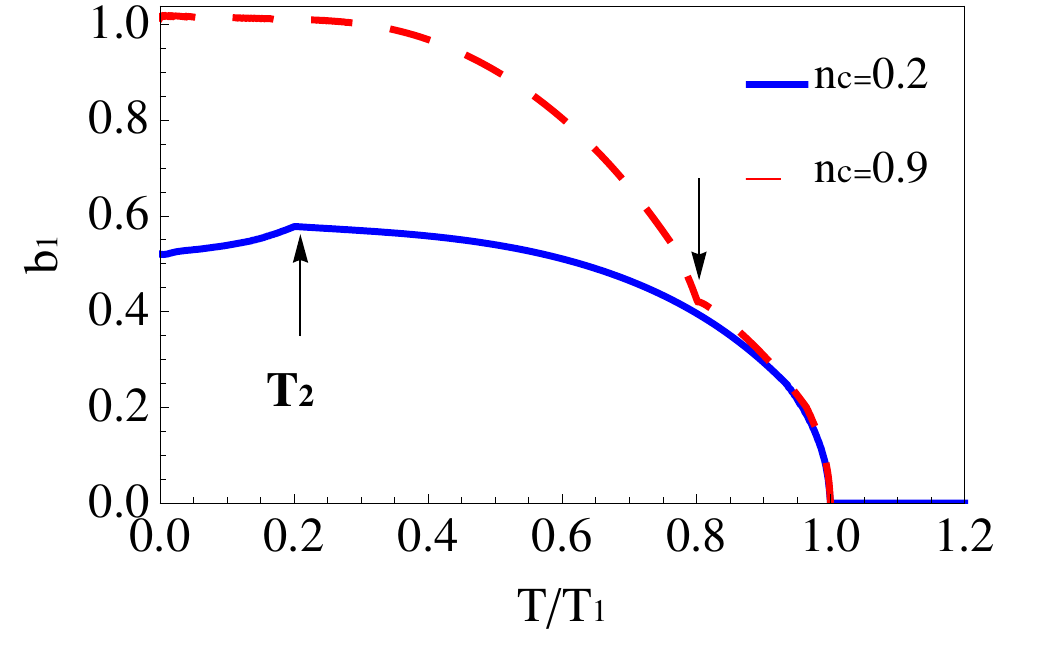}
\vspace{-0.4cm}
\caption{
Temperature dependence of the A-sublattice condensate $b_1$ for $J_2=7, J_1=8$ at
different filling numbers $n_c=0.2, 0.9$. The temperature scale is normalized by $T_1$.
The arrows point to the kinks occurring at the second Kondo scale $T_2$. The bare density
of states $\rho_0$ used is constant in its support with a half-bandwidth $D=10$.
}
\label{b1_D10_J27}
\end{figure}

So far, we have considered the effect of the A-sublattice Kondo screening on the
B-sublattice one, mainly in the extreme inhomogeneous case of small $J_2$. We
now discuss the ``feedback'' effect of the B-sublattice screening on the A-sublattice one. We
have already emphasized that the A-sublattice Kondo scale is the same as the one obtained
for the homogeneous case $J_1=J_2$. That is because this scale depends on the bare
density of states of $c$ electrons. Fig.~\ref{b1_D10_J27} shows the generic temperature
evolution of the A-sublattice condensate $b_1$ at two different filling numbers $n_c$. A kink is
observed as soon as the B-sublattice Kondo effect sets in. Below quarter-filling, when an
inhibition of the B-sublattice Kondo screening is observed, $b_1$ is reduced below $T_2$,
whereas it increases above quarter-filling, saturating at low temperatures. This means
that the two Kondo effects are either mutually co-operative (above quarter filling) or
mutually competitive (below quarter filling).
\begin{figure}[b]
\vspace{-0.cm}
\centering
 \includegraphics[width = 8.5cm]{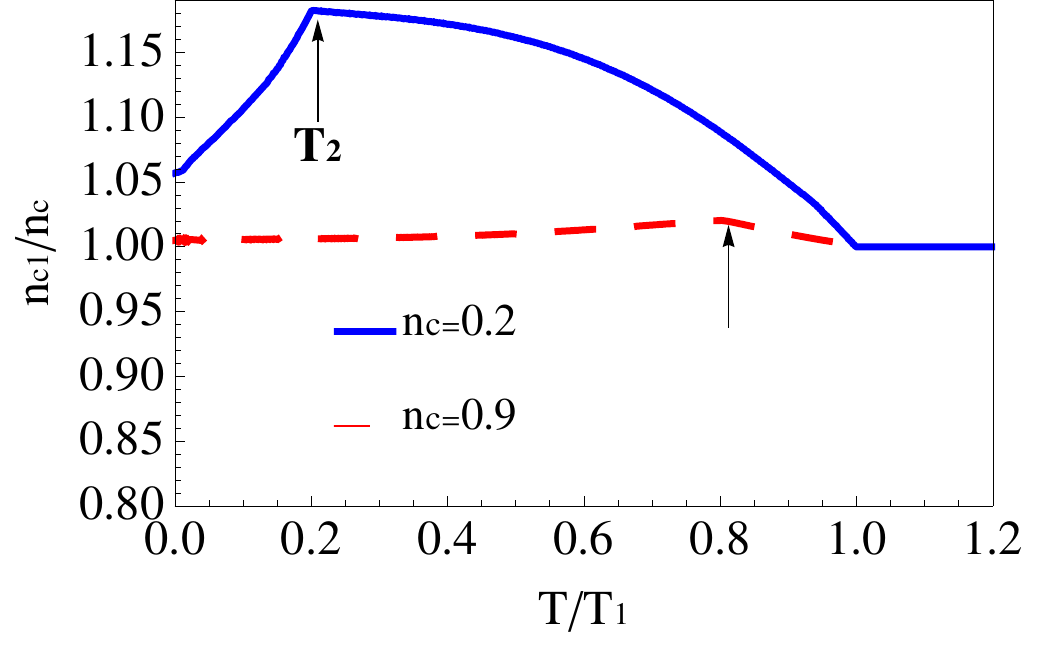}
\vspace{-0.4cm}
\caption{
Temperature evolution of the average occupation number $n_{c1}$ in the first sublattice
for $J_2=7, J_1=8$ at different filling numbers $n_c=0.2, 0.9$. The temperature scale is
normalized by $T_1$ and $n_{c1}$ is normalized by the total filling number $n_c$. The
arrows point to the kinks occurring at the second Kondo scale $T_2$. The bare density of
states $\rho_0$ used is constant in its support with a half-bandwidth $D=10$.
}
\label{Nc1_D10_J27}
\end{figure}
Fig.~\ref{Nc1_D10_J27} shows the generic temperature evolution of the average occupation
number $n_{c1}$ in the A sublattice at two different fillings $n_c=0.2, 0.9$. It turns
out that for all fillings $n_c$ this quantity increases as soon as the A-sublattice
Kondo effect sets in, i.e there are more electrons on the A sublattice as compared to B
(recall that, for $T>T_1$, both fillings are equal in our model). At the onset of the
B-sublattice Kondo effect, $n_{c1}$ reduces again without necessarily recovering its
high-temperature value.

The fact that  $n_{c1}$ increases for all values of filling number $n_c$ as soon as the
A-sublattice Kondo effect sets in means that the latter sustains itself not only by involving
low energy, i.e near the Fermi energy, conduction electrons in the second sublattice but
also \emph{high energy} ones. It shows also that the average number of $c$ electrons in the
B sublattice $n_{c2}$ plays little role, compared to the density of states
$\rho_{c2}$, in identifying a promotion or inhibition regime for the second Kondo effect.

\section{Low-temperature properties of the Fermi-liquid state}
\label{sec:FL}

At sufficiently low temperatures, we expect our model system to behave as a Fermi liquid
(neglecting possible magnetic or superconducting instabilities and provided that no Kondo
insulator obtains). The Fermi liquid can be characterized by its linear-in-temperature
specific heat $C(T)/T \approx \gamma$, constant magnetic susceptibility $\chi_0$ and
quadratic-in-temperature resistivity $\rho(T)-\rho_0 \approx A T^2$. This Fermi-liquid
behavior will be realized below a so-called coherence scale $\Tcoh$, which governs the
low-temperature properties of thermodynamic and transport quantities and takes the role
of an effective Fermi energy of the heavy quasiparticles.

For the case of a homogeneous Kondo lattice, studies using local self-energy
approximations to the Kondo lattice model (i.e. dynamical mean-field theory and slave
bosons) have shown that is different from $\TK$, but shares its exponential dependence on
the Kondo coupling $J$.\cite{burdin1,burdin2,pruschke,grenzebach} In particular, their ratio has
been found to be independent of $J$. In this section, we explore this issue in the case
of two inequivalent local moments per unit cell.  The effective four-band picture is
particularly appropriate to describe the Fermi-liquid regime that sets in at very low
temperatures.

In the low-temperature limit, thermodynamic quantities, such as the electronic entropy $S_e(T)$
or the electronic specific heat coefficient $C(T)$, are determined by the density of
states at the Fermi level $\rho(0)$
\beq
S_e(T\rightarrow 0) = \frac{\pi^2}{6}\rho(0) T, \,\, C(T \rightarrow 0)=\frac{\pi^2}{6}\rho(0)T.
\eeq
Following Ref.~\onlinecite{burdin1}, one may define a coherence scale in terms of
the zero-temperature value of the average renormalized density of electronic states at
the Fermi level $\rho(0)=\frac{1}{2}\sum_{\nu} \left ( \rho_{c \nu}(0)+\rho_{f \nu}(0)
\right )$ as follows (recall $k_B=1$)\cite{coherence_foot}
\beq
\bTcoh\equiv \frac{1}{4 \rho(0, T=0)}
\eeq
so that the low-temperature expansion of thermodynamic and transport quantities may be
written in terms of powers of $T/\bTcoh$. The same applies to our mean-field parameters which are expected to saturate to their zero-temperature value sufficiently below
$\bTcoh$. As $\bTcoh$ is, strictly speaking, a
scale characterizing the $T=0$ state only, we define $\Tcoh = {\rm min}\{\bTcoh,T_1,T_2\}$.
(Note that $\bTcoh < T_2$ except in the proximity to Kondo-insulating states.)

\subsection{Entropy}

\begin{figure}[t]
\centering
\includegraphics[width = 9.35cm]{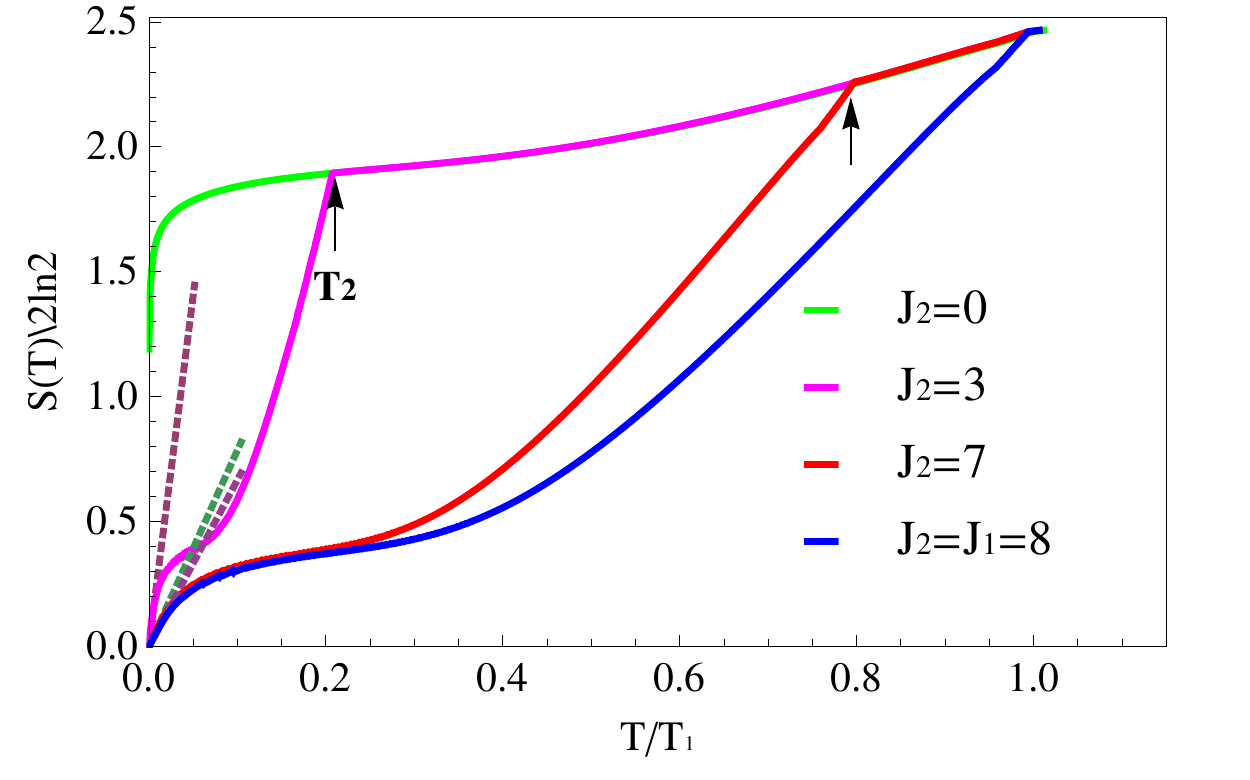}
\caption{
Total average entropy, normalized by $2\ln 2$,\cite{entropy_foot} at filling number
$n_c=0.9$ for different values of $J_2=0,3,7$ and $J_1=8$. The temperature scale is
normalized by $T_1$. Tiny dashed lines are linear fit at very low temperatures. The
arrows point to the kinks occurring at the second Kondo scale $T_2$. The bare density of
states $\rho_0$ used is constant in its support with a half-bandwidth $D=10$.
}
\label{ent_N09}
\end{figure}

Fig.~\ref{ent_N09} shows the temperature dependence of the total average entropy for different coupling
constants $J_2$ above quarter-filling, here $n_c=0.9$.

We first discuss the extremely inhomogeneous case $J_2=0$, corresponding to decoupled
$B$-sublattice moments. Here, the entropy starts to drop slowly below $T_1$, remaining
close to $4 \ln 2$, then decreases steeply towards low temperature to reach $2 \ln 2$ at
$T=0$. This residual entropy corresponds to the one of the unscreened local moments in
the second sublattice.\cite{entropy_foot}
Once this contribution is subtracted, a linear dependence is found at low temperature
consistent with the formation of a coherent Fermi liquid state, involving conduction
electrons in both sublattices as well as the screened local moments on the A sublattice
(while the B local moments are decoupled), with a finite ``coherence'' scale $\Tcoh$.

For finite values of $J_2$, a second kink signals the onset of the second Kondo screening
at $T_2$. Then a stronger decrease is observed in the entropy below this scale. This
decrease can be followed by a plateau-like feature close to half filling. (This feature
is absent below quarter-filling.) At lower temperatures, the entropy is again linear in
temperature signaling the emergence of a Fermi liquid regime out of the conduction
electrons and the local moments in both sublattices.

\subsection{Coherence scale}

\begin{figure}[t]
\vspace{-0.cm}
\centering
 \includegraphics[width = 9.cm, height=5.5cm]{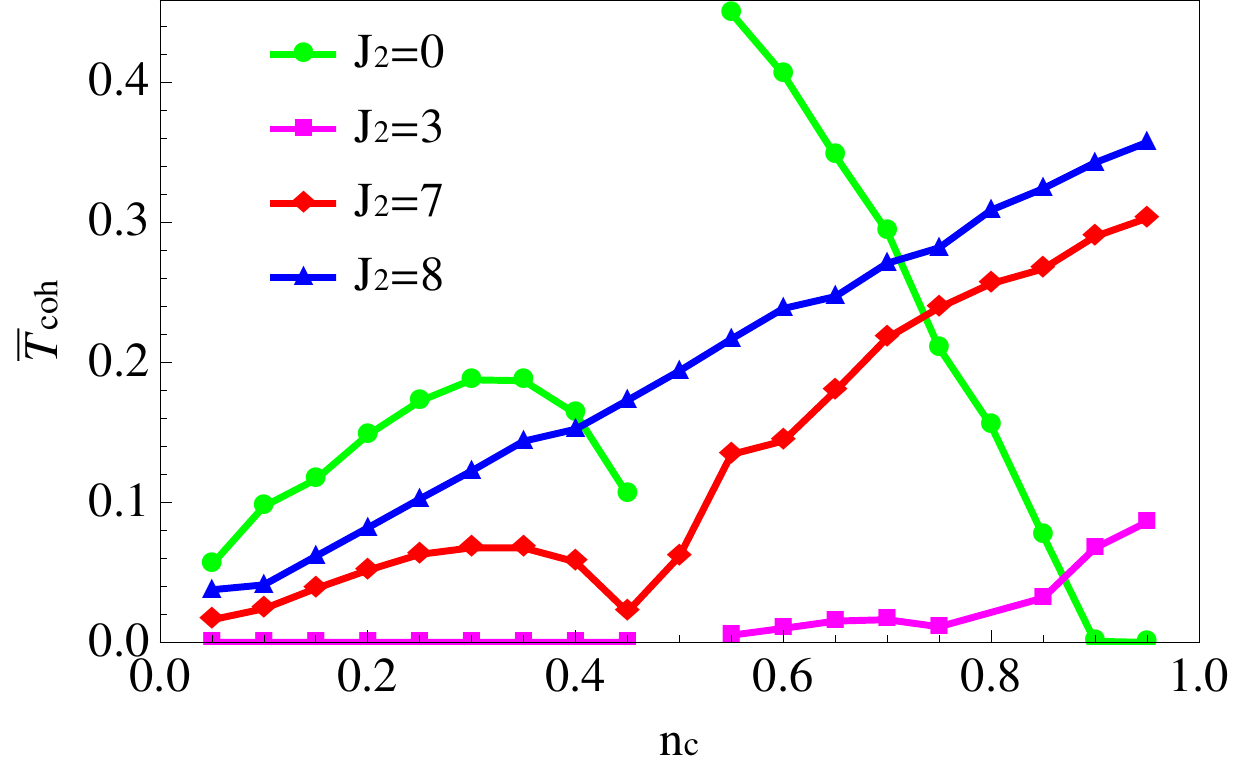}
\caption{
Evolution of the coherence scale $\bTcoh$ with the electronic filling $n_c$ for different
values of the coupling constant $J_2/J_1=0, 3/8, 7/8, 1$. The bare density of states
$\rho_0$ used is constant in its support with a half-bandwidth $D=10$.}
\label{T0_constant_D10}
\end{figure}

Fig.~\ref{T0_constant_D10} shows the evolution of the coherence scale $\bTcoh$ with the
filling number for different values of $J_2$. In all cases, $\bTcoh$ approaches zero as
$n_c\to 0$ as a result of ``exhaustion''.\cite{exhaus_foot}


Again, we start with the extremely inhomogeneous case $J_2=0$. Here, a band insulator is
obtained at quarter-filling and $\Tcoh$ is not defined.
Above quarter-filling, the coherence scale for $J_2=0$ decreases down to zero at half-filling.
This can be understood by looking at the renormalized band structure (see Fig.~
\ref{disp_D10_N06}). The width of the band crossing the Fermi level is of order $\mu$,
the chemical potential. As we increase the filling number $n_c$ towards half-filling,
this quantity, proportional to $\bTcoh$, decreases towards zero. The argument for $\bTcoh$
going to zero at half-filling relies on the fact that the bare DOS is symmetric, and thus
that $\mu=0$ at half-filling. However, we observed that the decrease of $\bTcoh$ above
quarter-filling holds even with the inclusion of next-nearest neighbor hopping, i.e. for
a asymmetric bare DOS, albeit not towards zero at half-filling.

\begin{figure}[t]
\centering
 \includegraphics[width = 9.cm]{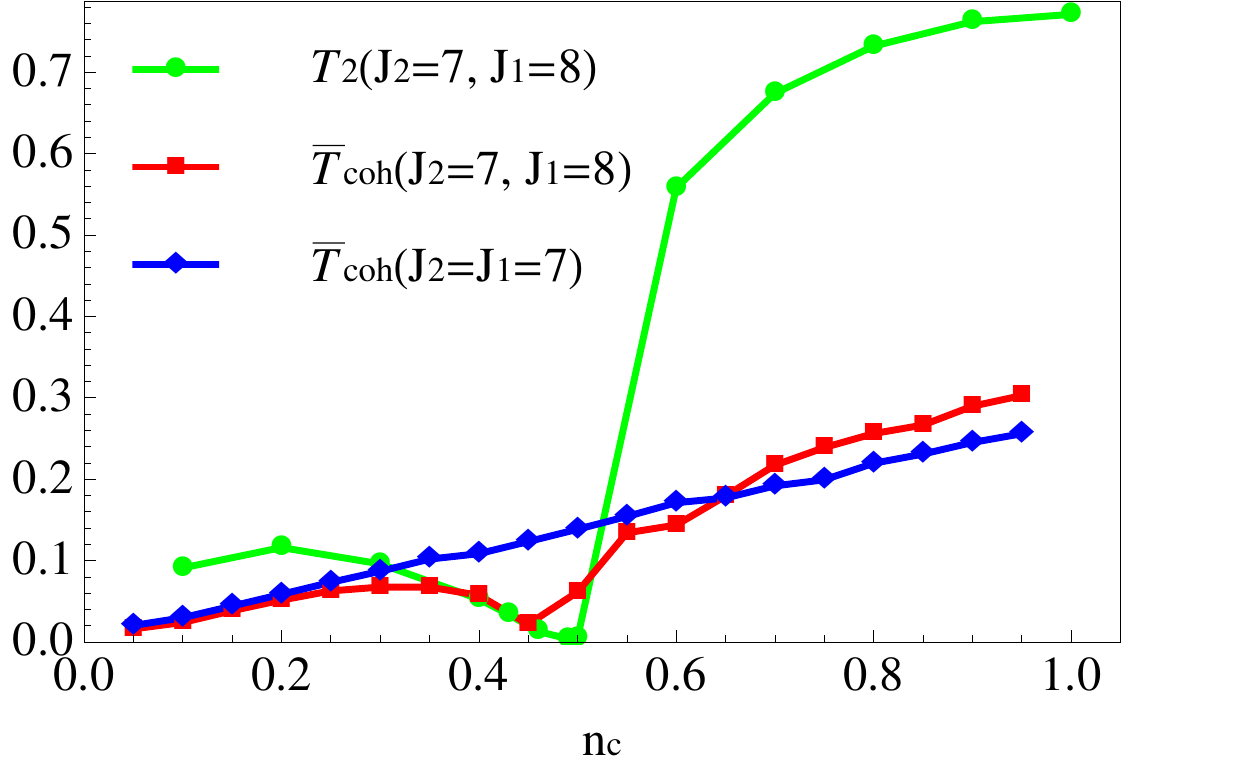}
 \includegraphics[width = 9.cm]{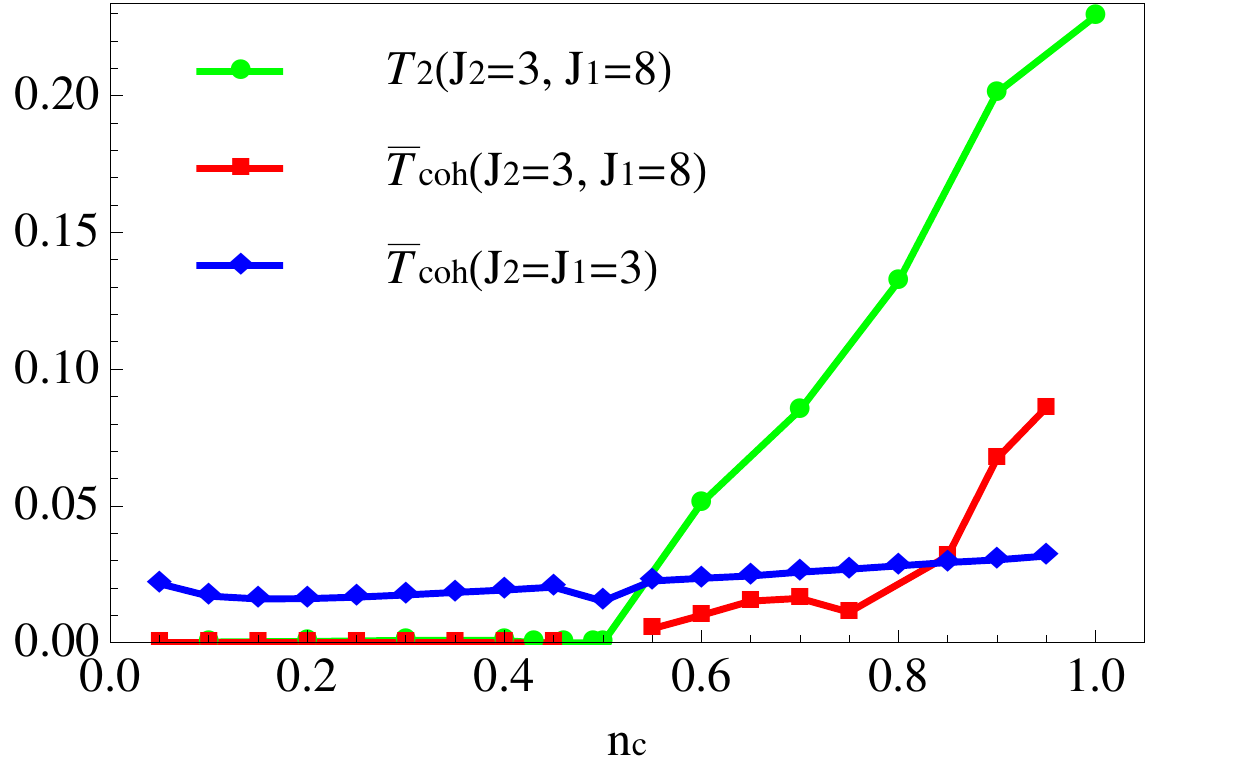}
\caption{
Evolution of the coherence scale $\bTcoh$ and the second Kondo scale $T_2$ with the
electronic filling $n_c$ for coupling constants $J_2/J_1=7/8$ (upper panel) and
$J_2/J_1=3/8$ (lower panel). Also shown is $\bTcoh$ for the homogeneous case $J_1=J_2$.
The bare density of states $\rho_0$ used is constant in
its support with a half-bandwidth $D=10$.
}
\label{various_T0_constant_D10}
\end{figure}

For finite $J_2$, the behavior of $\bTcoh$ is consistent with the observations regarding
the promotion or inhibition of the B-sublattice Kondo effect, as its scale $T_2$ sets an
upper bound on $\bTcoh$. This is more easily seen in Fig.~\ref{various_T0_constant_D10},
where we compare $\bTcoh$ to the homogeneous case's result. At low fillings, the
coherence scale $\bTcoh$ is smaller compared to the homogeneous case, consistent with the
inhibition of the second Kondo effect: the mean-field parameters saturate at lower
temperatures compared to the homogeneous case. The opposite happens above, where $\bTcoh$
is increased, consistent with the promotion of the second Kondo effect. Its effect on
$\bTcoh$ is most pronounced close to half filling.

It is worth discussing the contributions to the total DOS $\rho(0)$, which defines our
$\bTcoh$. Consider the ratio between the densities of states of the two
sublattices:
\bea
\frac{\rho_1(0)}{\rho_2(0)}&\equiv& \frac{(1+\frac{J_1 b_1^2}{\lambda_1^2}) }{(1+\frac{J_2 b_2^2}{\lambda_2^2}) } \frac{\rho_{c1}(0)}{\rho_{c2}(0)}\nn
&=&\frac{1}{\alpha(0)^2} \frac{(1+\frac{J_1 b_1^2}{\lambda_1^2}) }{(1+\frac{J_2 b_2^2}{\lambda_2^2})},
\label{ratio_rho12}
\eea
where $\alpha(0)$ is the boost factor given by
\beq
\alpha(0)=\sqrt{\frac{1-\frac{J_1 b_1^2}{\lambda_1 \mu}}{1-\frac{J_2 b_2^2}{\lambda_2 \mu}}},\nonumber
\eeq
see Eq.~\eqref{app:densities} in Appendix~\ref{app:spectral_densities} for details.
The ratio (\ref{ratio_rho12}) is mainly controlled by two quantities: the boost factor
$\alpha$ and the relative contributions coming from the heavy $f$-like bands forming below
the Kondo scales.

\begin{figure}[t]
\vspace{-0.cm}
\centering
\includegraphics[width = 9.cm]{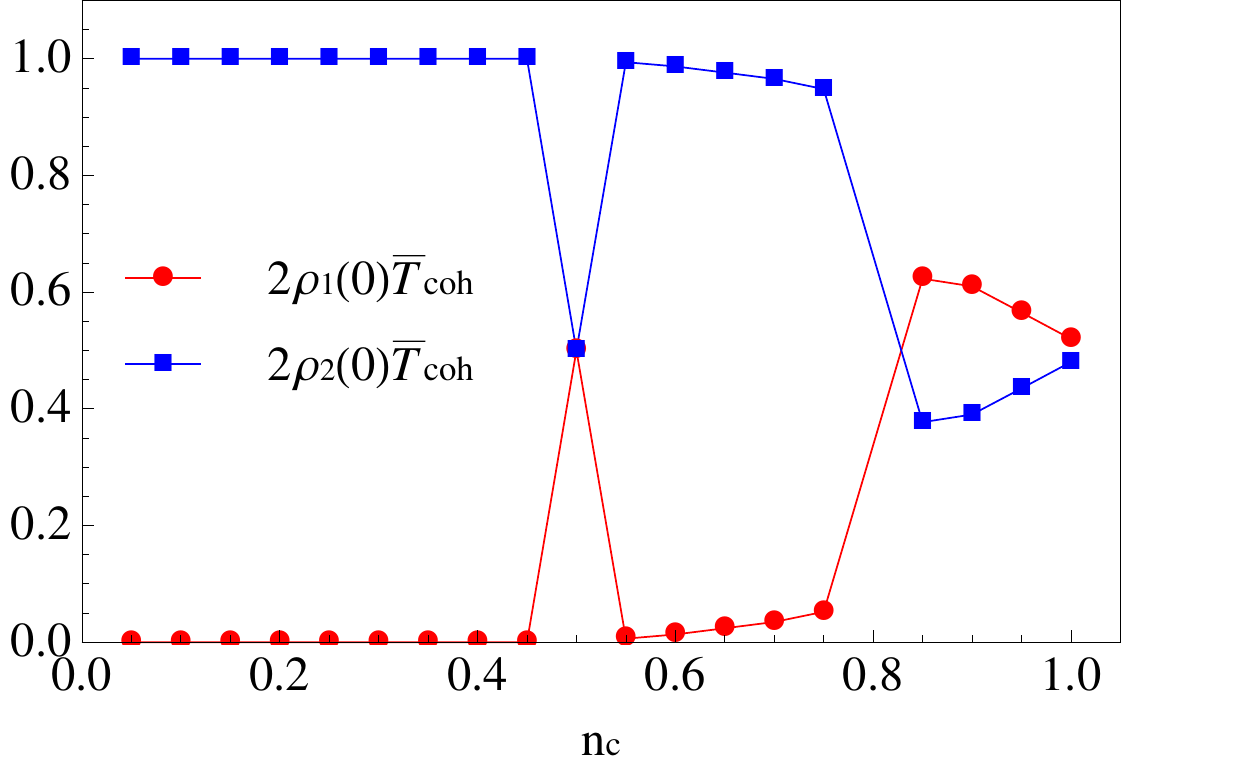}
\includegraphics[width = 9cm]{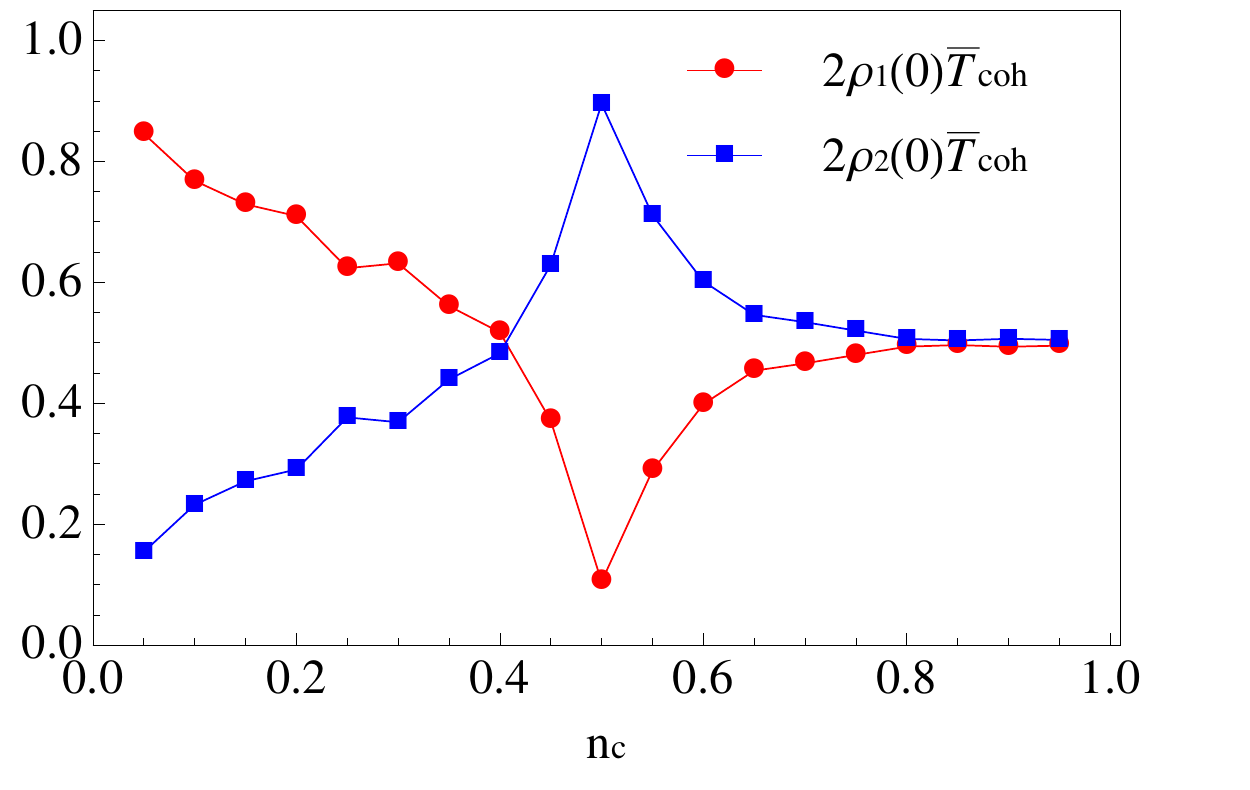}
\caption{
Contributions of each sublattice to the low-temperature average density of states at the
Fermi level $\rho(0)=1/4\bTcoh$, illustrated for different electronic filling $n_c$ and
$J_2/J_1=3/8$ (top) and $J_2/J_1=7/8$ (bottom). $\rho_\nu \equiv \rho_{c \nu} +
\rho_{f \nu}$ is the density of states of sublattice $\nu$. The bare density of states
$\rho_0$ used is constant in its support.
}
\label{T0T12_constant_D10J1J20}
\end{figure}

Fig.~(\ref{T0T12_constant_D10J1J20}) shows the contributions of each sublattice to the total DOS
$\rho(0)=1/(4\bTcoh)$ for different electronic fillings $n_c$ and couplings $J_2$.
For $0<J_2/J_1 \ll 1$, the total DOS is dominated mainly by contributions
from the B sublattice for the lowest electronic filling up to $n_c\sim 0.75$ then
decreases below the contribution from the A sublattice before converging to one half
towards half filling. In this regime of Kondo couplings, and for low electronic filling, the
main contribution comes from the $f$-like heavy band formed out of local moments
in the B sublattice (regardless of the small boost factor).
We note that the detailed behavior near $n_c\sim0.8$ depends on the input $c$-electron
DOS and is different for the square-lattice case.

For $J_2/J_1 \lesssim 1$, contributions from the A sublattice dominates for very low
electronic fillings before being over passed by the contribution from the B sublattice
already below quarter filling. The latter contribution attains its relative maximum at
quarter filling before reducing towards one half, both contributions being equal at half
filling. Here the effect of the boost factor wins over the effect originating from the
two $f$-like heavy bands and mainly sets the behavior of the ratio (\ref{ratio_rho12})

An interesting feature for the different investigated shapes of DOS and $0<J_2/J_1$
ratios, when only nearest-neighbor hopping is considered, is the convergence towards
equal contributions from each sublattice when approaching half filling, $n_c\to 1$. This
is due to the symmetry of the DOS that imposes $\mu=0$ at half-filling. In this case, one
can check from \eqref{ratio_rho12} that $\mu \rightarrow 0$ causes $\rho_1(0)/\rho_2(0)
\rightarrow 1$ for all values of $J_2$ and $J_1$.



\section{Discussion}
\label{sec:conclusions}

\subsection{Summary}

In this paper, we have investigated the interplay between two sublattice Kondo effects in
a bi-partite Kondo lattice model with two spin-$1/2$ local moments per unit cell, with
Kondo couplings $J_1>J_2$ on sublattices A,B. We have found that the conduction-band
filling $n_c$ determines how the B-sublattice Kondo effect is affected by the onset of
A-sublattice Kondo screening: it is generically promoted for $n_c>0.5$ and inhibited for
$n_c<0.5$. Our collected results thus show that quarter filling, $n_c=0.5$, is the
boundary separating regimes where the two Kondo effects are either mutually co-operative
($n_c>1/2$) or mutually competitive ($n_c<1/2$). A strong competition in the latter
regime implies that {\em small} differences between $J_1$ and $J_2$ induce {\em large}
differences between $T_1$ and $T_2$ as a result of non-trivial feedback effects, see
Fig.~\ref{Tk_J_D10}a.
Therefore, the regime of partial screening advocated in the introduction may be reached
easily for small conduction band filling.

\subsection{Relation to depleted Kondo lattices}

Heavy fermion systems with randomly depleted local moments constitute an interesting
route to interpolate between the physics of isolated Kondo impurities and coherent
Kondo lattice behavior. Experimentally, depletion is realized by substituting magnetic by
non-magnetic ions, as e.g. in Ce$_{1-x}$La$_x$Al$_3$. Theoretically, a number of papers
have treated depleted Kondo lattices,\cite{kaul_dep,burdin_dep,ogata_dep} with randomly
distributed local moments with concentration $n_f<1$, mainly using local self-energy
approximations, combined either with CPA or with a numerically exact treatment of spatial
disorder. One of the remarkable findings is that $n_f = n_c$ marks a boundary between two
distinct regimes: For $n_f < n_c$ the behavior is akin to that of dilute impurities,
i.e., lattice effects appear to be small, whereas $n_f > n_c$ leads to Kondo lattice
behavior. Although both regimes are Fermi-liquid-like at lowest temperatures, the
crossover is rather pronounced e.g. in the coherence temperature $\Tcoh(n_f)$ which is
approximately constant for $n_f<n_c$, but rapidly drops upon increasing $n_f$ beyond
$n_c$. The latter behavior may be related to ``exhaustion'', i.e., the lack of conduction
electrons to screen the moments. It has been speculated that the boundary $n_f=n_c$ is
inherited from the strong-coupling limit, $J\to\infty$, where $n_f=n_c$ represents a
(random) Kondo insulator.\cite{kaul_dep}

We believe that the two seemingly different problems of depleted Kondo lattices and Kondo
lattices with inequivalent local moments are related in the following way:
For $T_2<T<T_1$ in the latter problem, Kondo screening is ``active'' for half of the
moments. This can be read as an $n_f=1/2$ depleted Kondo lattice, albeit without
randomness.
Carrying over the results of Refs.~\onlinecite{kaul_dep,burdin_dep,ogata_dep} suggests
that $n_c<1/2$ then corresponds to a collective regime with exhaustion playing a role. As
a result, conduction electrons are strongly involved in screening A-sublattice moments, such
that B-sublattice screening is suppressed and the two Kondo effects compete. In contrast, for
$n_c>1/2$ a single-ion picture is more appropriate: Kondo screening of A sites injects
states near the Fermi level, such that B-sublattice screening is promoted and the two Kondo
effects co-operate. These arguments, although somewhat simplistic, are consistent with
our numerical results for $\rho_{c2}$ in Sec.~\ref{sec:fill}.

\subsection{Extensions}

A natural extension of the calculations presented here is to take into account non-local
magnetic correlations, due to either RKKY interactions or a direct Heisenberg-like
exchange. In addition to the standard Kondo lattice phenomenology, i.e., the occurrence
of magnetic (or spin-liquid) phases for dominant inter-moment exchange, interesting
physics can be expected in the regime of partial screening, see Fig.~\ref{fig:pd}.
For instance, for small $J_2$ the B-sublattice moments are more susceptible to non-local
effects than the A-sublattice moments, paving the way to either non-local
B-sublattice ``screening'' or B-sublattice-dominated magnetic order.

One might also speculate about variants of the Kondo-breakdown (KB) scenario for quantum
criticality, originally proposed as a possible description of some fascinating non-Fermi
liquid properties in heavy fermion systems.\cite{si,senthil,paul} A central role in this
scenario is played by so-called fractionalized Fermi-liquid states, formed out of
conduction electrons weakly coupled to an exotic (i.e. fractionalized) spin
liquid.\cite{senthil}
In the present situation with two local moments per unit cell, a ``conventional'' spin
liquid with intra-cell dimer formation are possible, leading to a light metallic state
with Fermi-liquid properties, similar to that in the bilayer Kondo lattice model of
Ref.~\onlinecite{senthil05}. However, exotic spin liquids may be formed e.g. from the
B-sublattice moments only, such that a fractionalized Fermi liquid may be realized in the
regime of partial screening.
A comprehensive study of these issues is left for future work.

\subsection{Experiments}

At present, relatively little detailed information is available about the low-temperature
properties of the (RE)$_3$Pd$_{20}$X$_6$ compounds mentioned in the introduction.
Clearly, a detailed characterization of the ordered phases, e.g., by neutron scattering,
is desirable to put those in the global perspective advocated here.

Based on our results, we propose to perform systematic doping studies of these compounds,
using dopants of different valence. As shown here, variations in the conduction electron
filling can be expected to significantly affect the different Kondo scales in the system.


\begin{acknowledgments}
We thank A. Hackl, A. Mitchell, A. Rosch, and E. Sela for fruitful discussions.
This research was supported by the DFG through SFB 608 and FOR 960.
\end{acknowledgments}

\appendix

\section{Review of slave-boson results for the standard Kondo lattice}
\label{app:Kondo}

Here, we recall some results from the mean-field approximation of the Kondo lattice
problem. An extensive presentation can be found in Ref.~\onlinecite{burdin2}.

The mean-field equations are written as follows
\bea
b&=&\sqrt{J}\int_{-\infty}^{\infty}d \omega \,\, \rho_{fc}(\omega)n_F(\omega),\nn
1&=&2\int_{-\infty}^{\infty}d \omega \,\, \rho_{f}(\omega)n_F(\omega), \\
n_c&=&2\int_{-\infty}^{\infty}d \omega \,\, \rho_{c}(\omega)n_F(\omega),\nonumber
\label{app:MFE}
\eea
where $n_F(\omega)$ is the Fermi function and $\rho_{c}, \rho_{f}, \rho_{fc}$ are the
spectral densities of the local single-particle Green's functions given as
\bea
\rho_{c}(\omega)&\equiv&-\frac{1}{\pi} \Im{\mathcal G_{cc}^+(\omega)}=\rho_{0}\left ( \epsilon(\omega)\right ),\nn
\rho_{fc}(\omega)&\equiv&-\frac{1}{\pi} \Im{\mathcal G_{fc}^+(\omega)}=\frac{\sqrt{J}b}{\omega +\lambda}\rho_{c}(\omega),\\
\rho_{f}(\omega)&\equiv&-\frac{1}{\pi} \Im{\mathcal G_{ff}^+(\omega)}=\frac{J b^2}{(\omega +\lambda)^2}\rho_{c}(\omega),
\nonumber
\eea
where $\mathcal G^+ (\omega)\equiv \mathcal G(i\omega \rightarrow \omega +i 0^+)$ is the retarded Green's function matrix and
$
\epsilon(\omega) = \omega+\mu-\frac{J b^2}{\omega +\lambda}.
$

Notice that
\beq
(n_c+1)\,{\rm mod}\,2 = \int_{-\infty}^{\mu_L}\rho_0(\omega),
\label{mulint}
\eeq
where $\mu_L\equiv \mu-\frac{J b^2}{\lambda}$ can be understood as a chemical potential
corresponding to the enlarged Fermi volume, which encompasses $n=n_c+1$ electrons when
$b\neq0$.

The Kondo temperature $\TK$ is defined as the temperature for which the effective
hybridization $\propto b$ vanishes. It is obtained from the mean-field equations
(\ref{app:MFE}) in the limit $b, \lambda \rightarrow 0$. This gives
\beq
\frac{2}{J}=\int_{-\infty}^{+\infty}d \omega \,\, \frac{\rho_0(\omega +\mu)}{\omega} \tanh{\left ( \frac{\omega}{2 \TK} \right )}.
\label{app:TK}
\eeq
For $\rho_0(\mu)\neq 0$, the main contribution to the integral in (\ref{app:TK}) comes
from $\omega=0$ and any finite coupling $J$ leads to $\TK > 0$. In the weak coupling
regime $J \ll D$, one finds \cite{burdin1}
\beq
\TK =\alpha_0 \sqrt{ D^2-\mu^2} F_K e^{-1/J\rho_o(\mu)},
\label{app:TK_weak}
\eeq
 where $\alpha_0=1.13$ is a numerical constant and
\beq
F_K=\exp{\left [ \int_{-(D+\mu)}^{D-\mu}d \omega \,\,\frac{\rho_0(\omega + \mu)-\rho_0(\mu)}{2|\omega|\rho_0(\mu)}\right ]}
\eeq
In the Kondo lattice problem, the Kondo scale $\TK$ is not necessarily the one that
characterizes the behavior of the low-temperature regime. Generally, the thermodynamic
quantities like the specific heat coefficient or the zero temperature magnetic
susceptibility are characterized by a second energy scale  given by \cite{burdin2}
\beq
\bTcoh = \frac{D+\mu}{\Delta
\mu}\frac{F_0}{\rho_0{\mu}}e^{-1/J\rho_o(\mu)}, \label{app:Tcohweak}
\eeq
where $\Delta\mu=\mu_L-\mu$ and
\beq
F_0=\exp{\left [ \int_{-(D+\mu)}^{\Delta \mu}d \omega \,\,\frac{\rho_0(\omega +
\mu)-\rho_0(\mu)}{|\omega|\rho_0(\mu)}\right ]}.
\eeq


From (\ref{app:TK_weak}) and (\ref{app:Tcohweak}), we see that the scales $\TK$ and $\bTcoh$
have the same exponential dependence on the coupling constant $J$. They can differ
considerably in their prefactors. Their ratio depends generally on the electronic
filling and the shape of the density of states. Well below $\Tcoh = \mbox{min} \{ \bTcoh, \TK \}$,
the system behaves as a Fermi liquid for less than half-filling.

\section{Results for the inhomogeneous square lattice}
\label{app:square_lattice}

\begin{figure}[t]
\centering
\begin{tabular}{ll}
\hspace{-0.5cm}\includegraphics[width = 4.65cm, height=3.5cm]{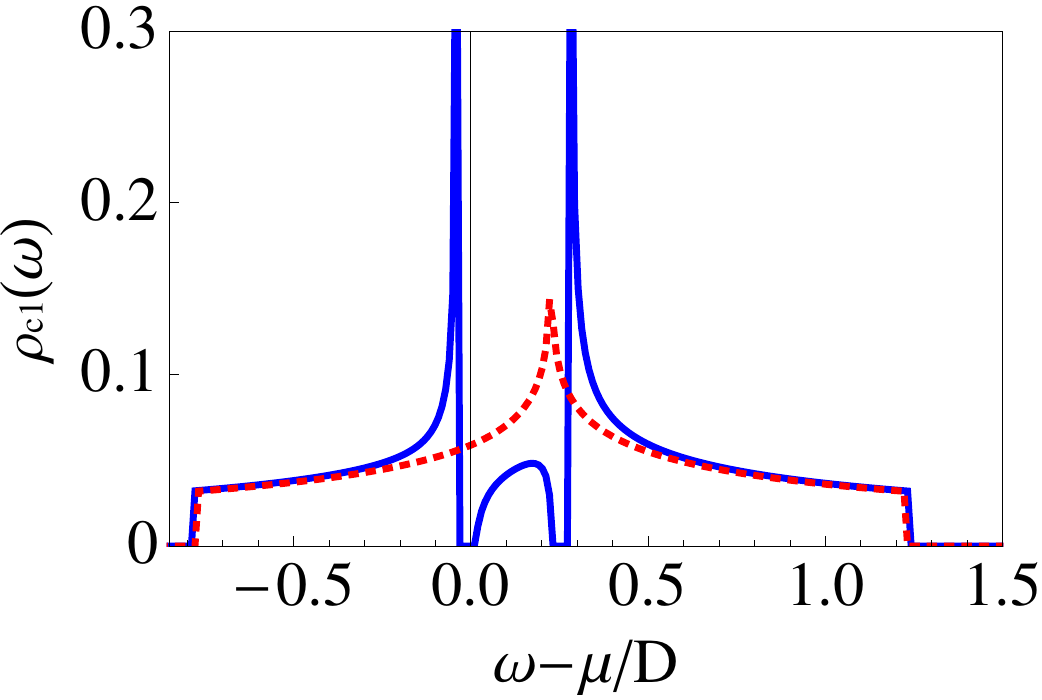}& \hspace{-0.2cm} \includegraphics[width = 4.65cm, height=3.5cm]{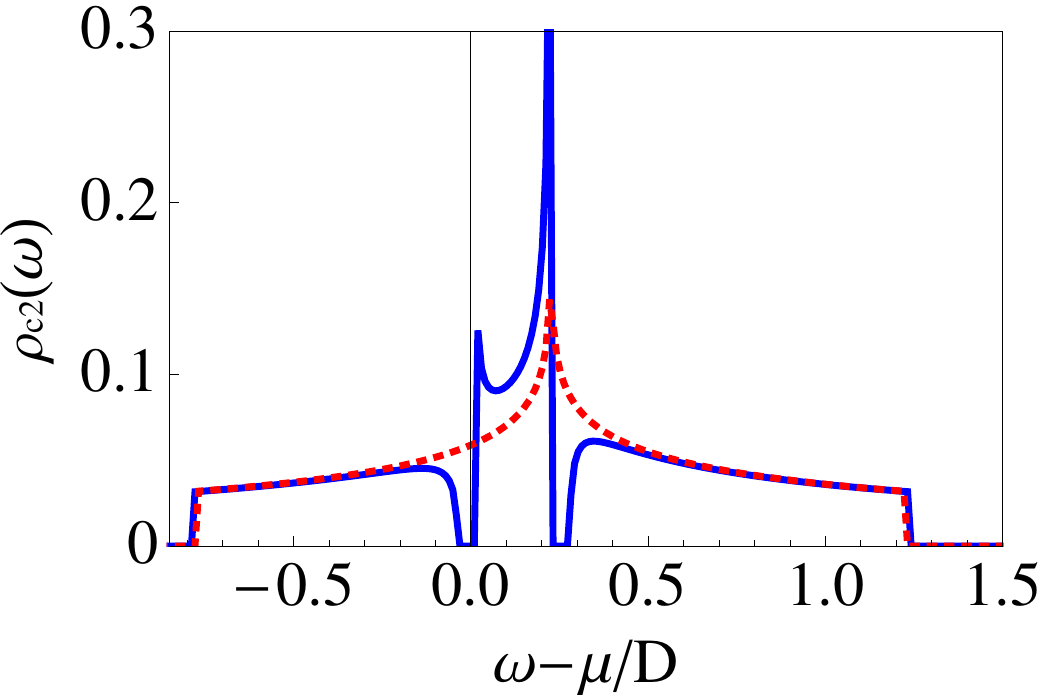}\\
\hspace{-0.5cm}\includegraphics[width = 4.65cm, height=3.5cm]{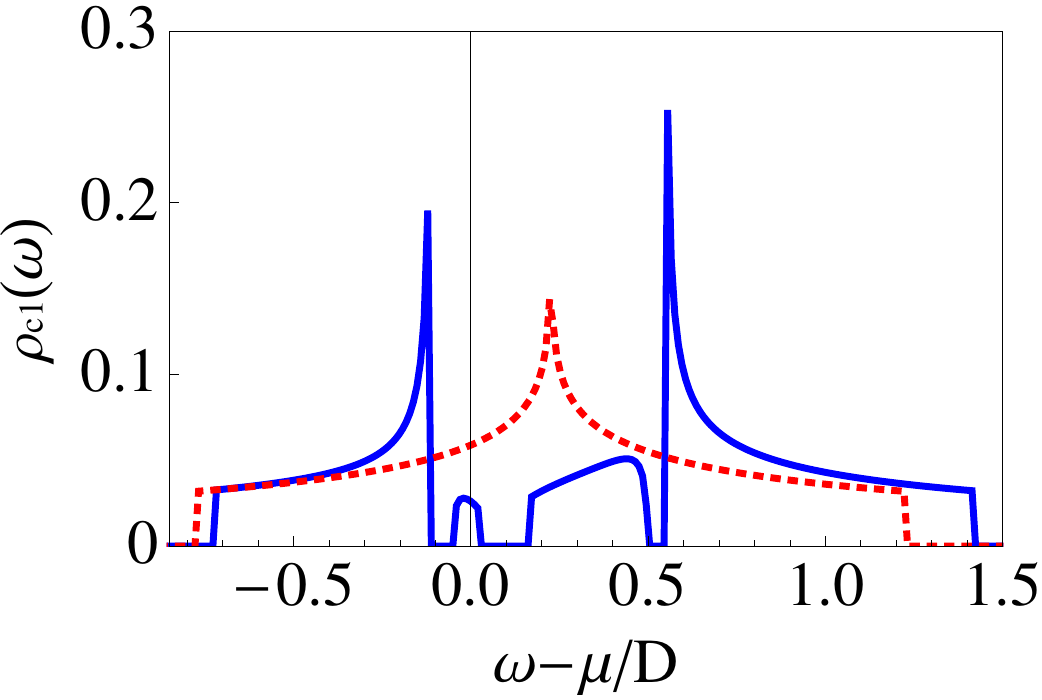}&\hspace{-0.2cm} \includegraphics[width = 4.65cm, height=3.5cm]{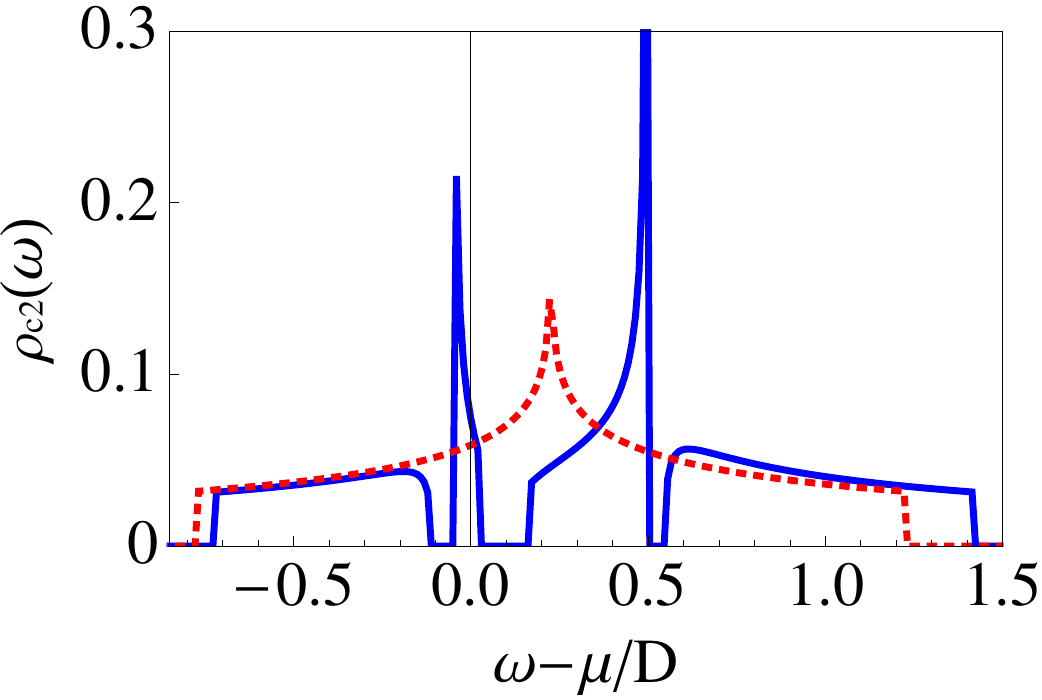}
\end{tabular}
\caption{
Densities of states of the c-electrons in each sublattice, left for sublattice A and
right for B, for $T_2<T<T_1$ (top panels) and $T<T_2$ (bottom panels) in the square
lattice case with no next-nearest neighbor hopping. The bare density of states for $T>
T_1$ is shown in dashed line in each panel. The half bandwidth is $D=10$.
}
\label{dens_square_D10_N06}
\end{figure}

\begin{figure}[b]
\centering
 \includegraphics[width = 8.cm]{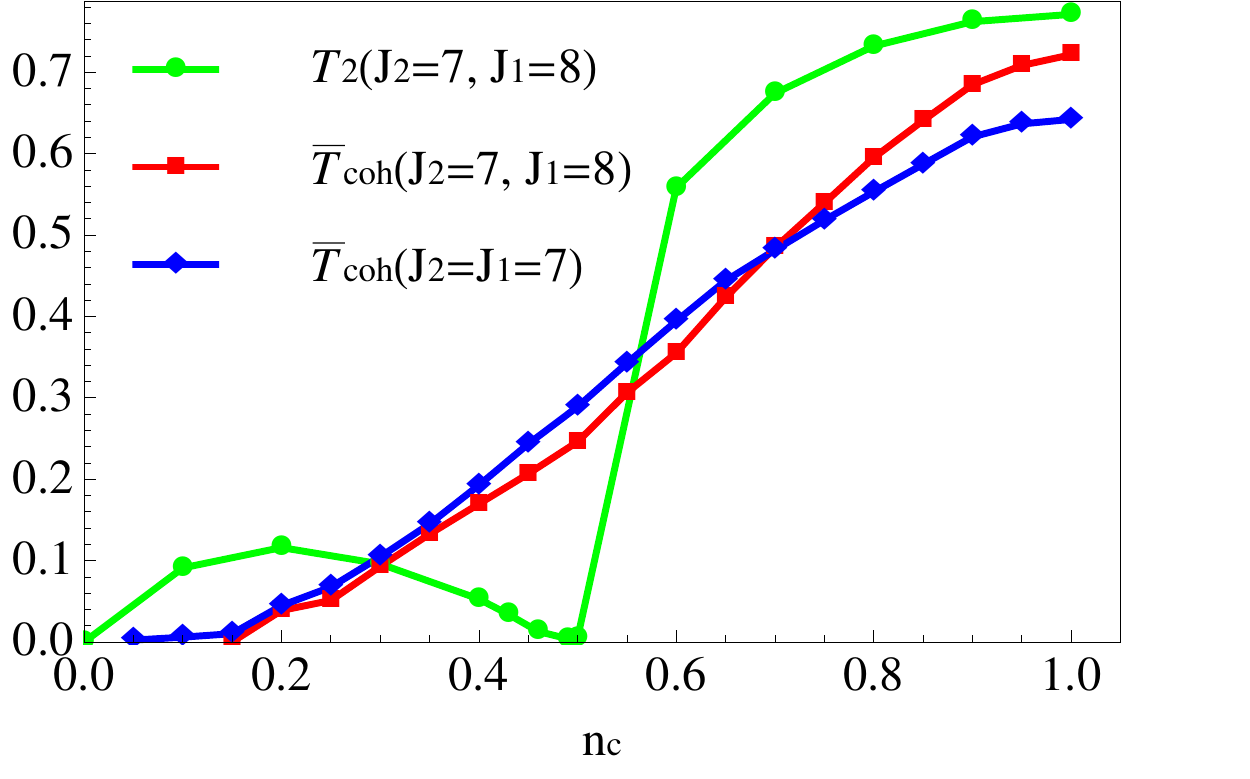}
\caption{
Evolution of the coherence scale $\bTcoh$ and the Kondo scale $T_2$ with the electronic
filling $n_c$ for a coupling constant $J_2/J_1=7/8$ and comparison with the standard
Kondo lattice  result for the square lattice case with no next-nearest neighbor hopping.
The half bandwidth is $D=10$. } \label{T0_square_D10J27}
\end{figure}

In this appendix, we present some results for the Kondo lattice problem with two
inequivalent local moments, Eq.~\eqref{model}, on a square lattice of $c$-electron orbitals.
Restricting ourselves first to the case of nearest-neighbor hopping only, we can apply
the numerical machinery as in the body of the paper, i.e., the momentum summations can be
converted into frequency integrals. To this end, one needs the $c$-electron density of
states, which is given by
\beq
\rho_\square(\omega)=\frac{2}{\pi^2 D}\mathcal{K}\left (\sqrt{1-\frac{\omega}{D}} \right )
\theta(D^2-\omega^2),
\eeq
where $\mathcal{K}(x)$ is the elliptic integral of the first kind, and $D=4t$ where $t$
is the nearest-neighbor hopping. This DOS is smooth and continuous, except at the
step-like band edges and at the $\omega=0$ van Hove singularity, where the DOS diverges
as $\rho_\square(\omega) \sim \frac{2}{\pi^2 D}\log \frac{4 D}{|\omega|}$.

Fig.~\ref{dens_square_D10_N06} shows the temperature evolution of the local DOS for the
$c$ electrons in each sublattice for the square lattice with no next-nearest neighbor hopping.
This evolution is consistent with Fig.~\ref{disp_D10_N06} showing the renormalized
dispersions at $T_2<T<T_1$ and $T<T_2$.

Fig.~\ref{T0_square_D10J27} shows the dependence of the coherence and Kondo scales on the
electronic filling number $n_c$ for $J_2 \simeq J_1$ and $J_2=J_1$ and how they compare
to the coherence scale in the homogeneous case. This is qualitatively similar to
Fig.~\ref{T0_constant_D10}: for low fillings, the coherence scale $\bTcoh$ is smaller
compared to the homogeneous Kondo lattice case then bigger above quarter filling, when we
approach half filling. Thus, the qualitative picture developed in the body of the paper
is reproduced for the square lattice.




\begin{figure}[t]
\centering
\begin{tabular}{ll}
\includegraphics[width = 5.25cm]{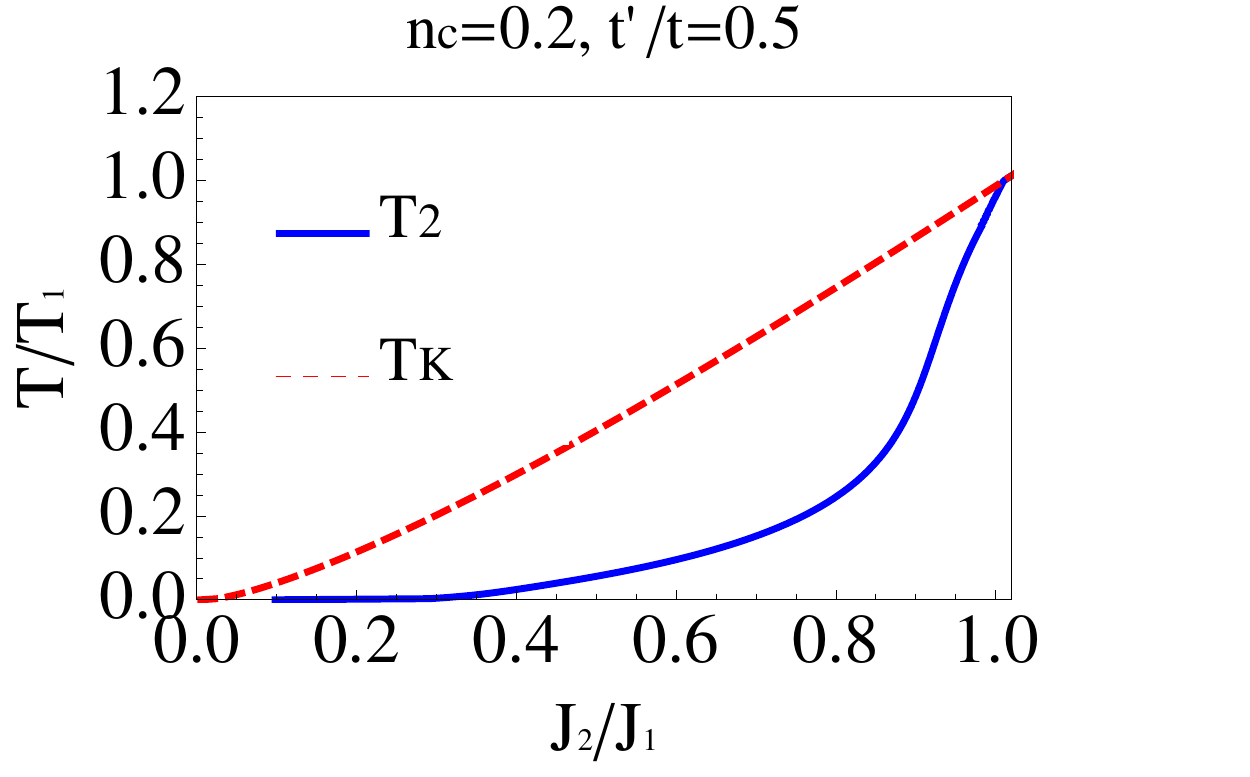}&
\hspace{-0.9cm}\includegraphics[width=5.25cm]{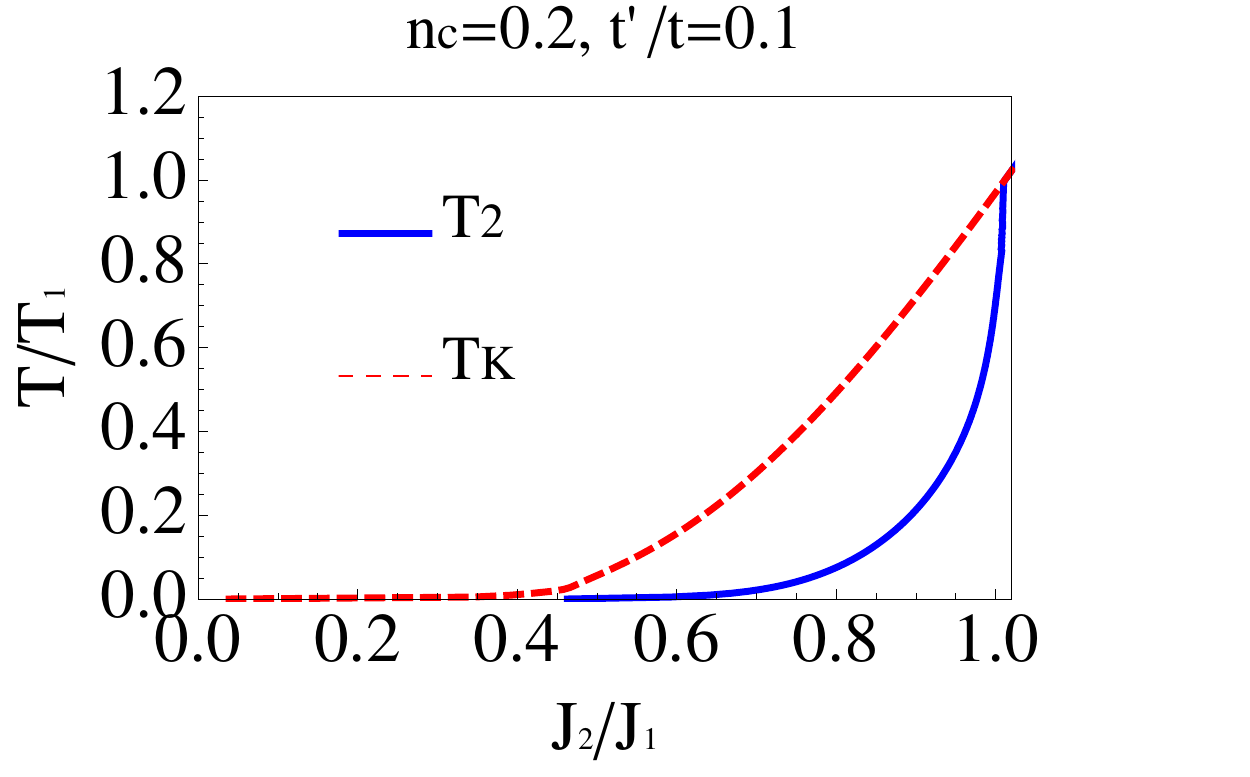}\\
\includegraphics[width = 5.25cm]{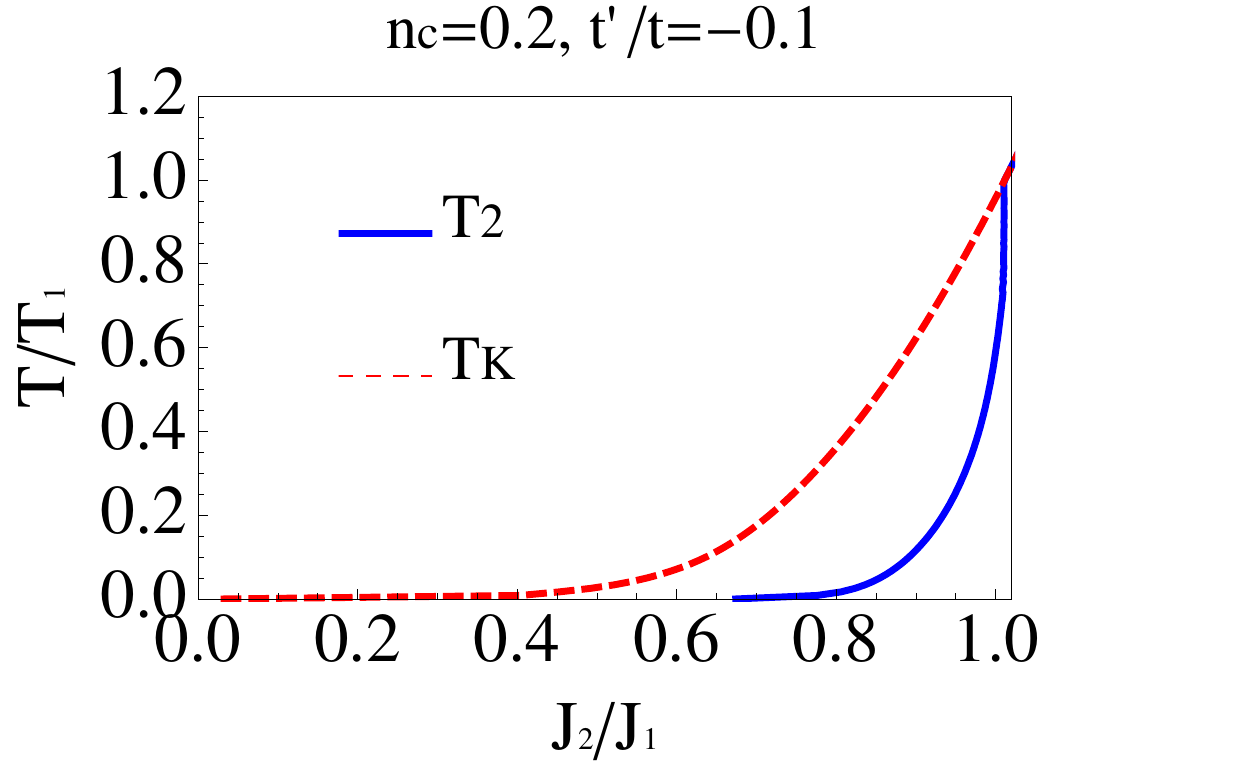}&
\hspace{-0.9cm}\includegraphics[width = 5.25cm]{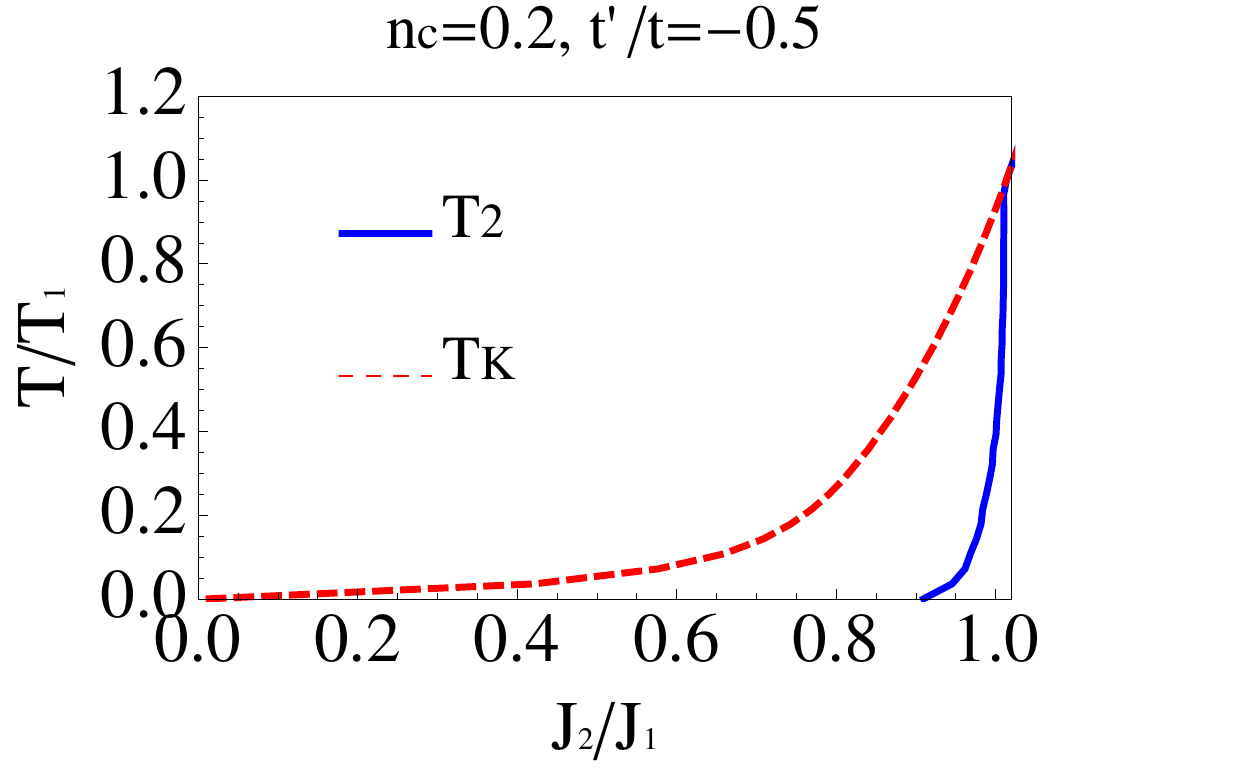}
\end{tabular}
\caption{
Evolution of the temperature scales $T_2(J_2)$ and $T_K(J_2)$, where $T_K(J_2)$ is the
Kondo scale obtained in the homogeneous Kondo lattice with a Kondo coupling $J_2$, with
the strength of the second Kondo coupling $J_2\leq J_1$ for an electronic filling
$n_c=0.2$ and various values of the next-nearest hopping parameter $t'$.
}
\label{square_tp_nc02_D10}
\end{figure}

\begin{figure}[t]
\centering
\begin{tabular}{ll}
\includegraphics[width = 5.25cm]{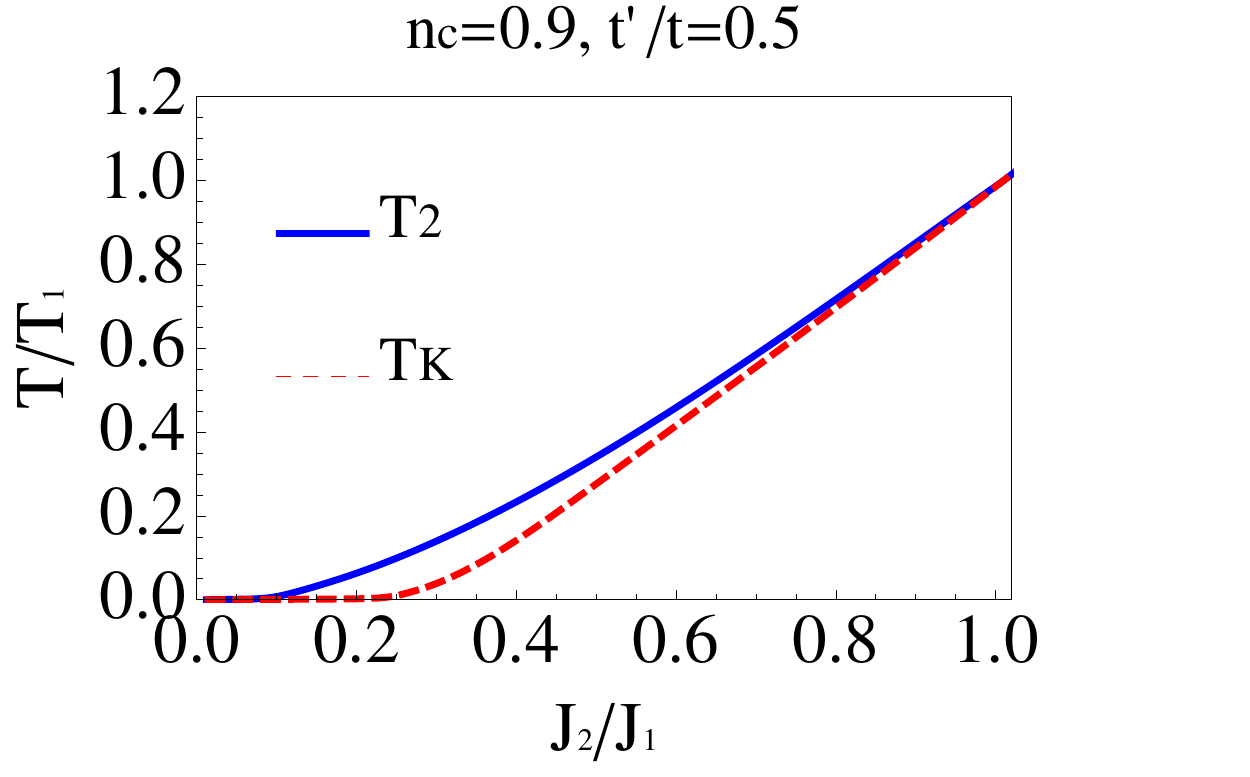}&
\hspace{-0.9cm}\includegraphics[width=5.25cm]{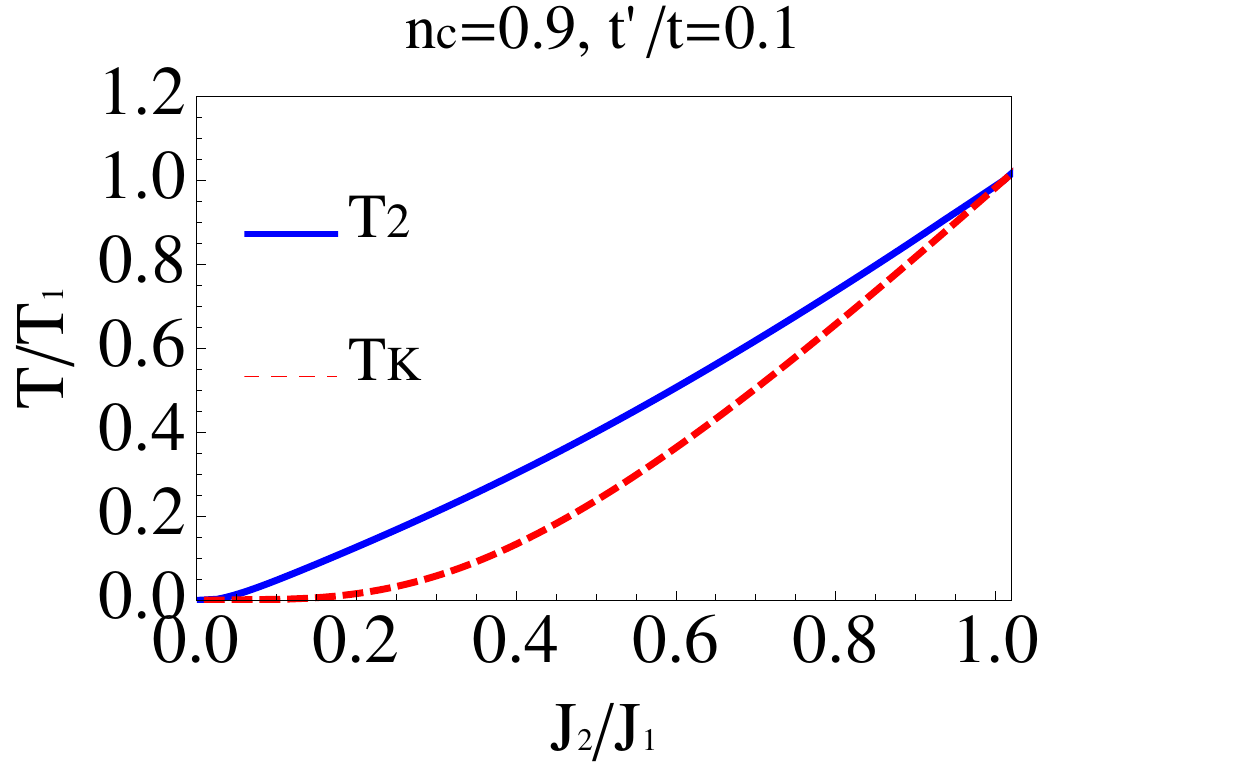}\\
\includegraphics[width = 5.25cm]{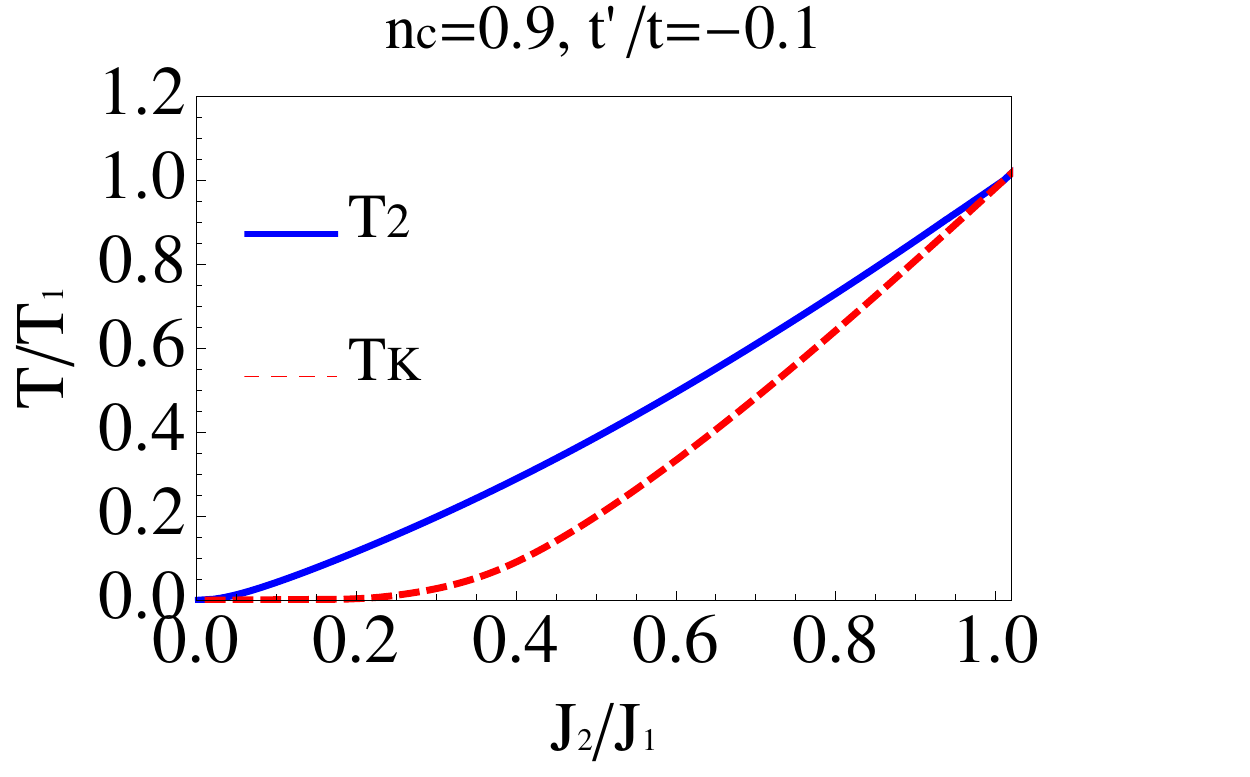}&
\hspace{-0.9cm}\includegraphics[width = 5.25cm]{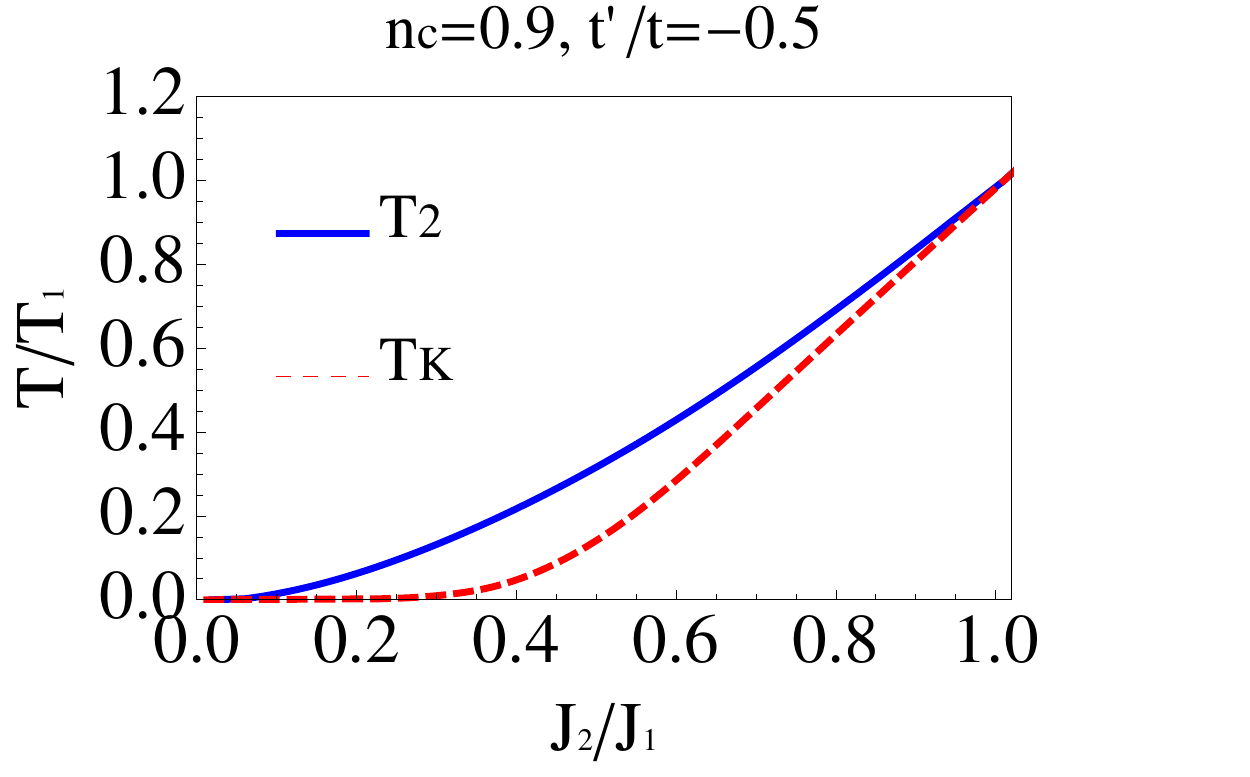}
\end{tabular}
\caption{
Evolution of the temperature scales $T_2(J_2)$ and $T_K(J_2)$, where $T_K(J_2)$ is the
Kondo scale obtained in the homogeneous Kondo lattice with a Kondo coupling $J_2$, with
the strength of the second Kondo coupling $J_2\leq J_1$ for an electronic filling
$n_c=0.9$ and various values of the next-nearest hopping parameter $t'$.
}
\label{square_tp_nc09_D10}
\end{figure}

Finally, we investigate the effect of the inclusion of next-nearest hopping, with a
parameter $t'$. Non-zero $t'$ results in non-vanishing momentum-dependent diagonal terms
$\xi_\bk=\zeta_\bk= - 4t' \cos{(k_x)}\cos{(k_y)}$ in the kinetic matrix $\mathcal E_\bk$
in \eqref{Green_matrix}. Then, momentum summations in the mean-field equations
\eqref{MFE} can no longer be turned into an integration over energies. Instead, the
mean-field equations are now solved on a discrete lattice of $64^2$ sites.

Figs.~\ref{square_tp_nc02_D10} and \ref{square_tp_nc09_D10} show the dependence of
the B-sublattice Kondo scale and the Kondo scale in the homogeneous case on the electronic
filling $n_c$ for various values of the next-nearest hopping parameter $t'$.

Our conclusion on the promotion or the inhibition of the second Kondo effect is found to
be robust against inclusion of next-nearest neighbor hopping, namely inhibition below quarter-filling
and promotion above quarter-filling.


\section{Spectral densities}
\label{app:spectral_densities}

The mean-field equations \eqref{MFE} can be written alternatively using the spectral
densities of the local single-particle Green's functions
%
\bea
b_\nu &=&  \sqrt{J_\nu} \int_{-\infty}^{\infty} d\omega\,\, \rho_{fc \nu}(\omega)n_F(\omega), \nn
\label{app:MFdensity}
1&=&2\int_{-\infty}^{\infty} d\omega\,\, \rho_{f\nu}(\omega)n_F(\omega),\\
n_{c\nu}&=&2\int_{-\infty}^{\infty} d\omega\,\, \rho_{c\nu}(\omega)n_F(\omega),\nonumber
\eea
where $n_F(\omega)=1/(1+e^{\omega/T})$ is the Fermi function. If we consider only nearest-neighbor hopping, we can have analytic expressions for the spectral densities $\rho_{c\nu}, \rho_{f\nu}, \rho_{fc\nu}$
\bea
\rho_{c\nu}(\omega)&=&\alpha^\nu(\omega)\rho_0 \left (\epsilon(\omega) \right)\,\,, \nn
\label{app:densities}
\rho_{f\nu}(\omega)&=&\frac{J_\nu b_\nu^2}{(\omega+\lambda_\nu)^2}\rho_{c\nu}(\omega)\,\,, \\
\rho^\nu_{fc}(\omega)&=&\frac{\sqrt{J_\nu}b_\nu}{\omega+\lambda_\nu}\rho_{c\nu}(\omega)\,\,,\nonumber
\eea
where
\beq
\alpha^\nu(\omega)=\sum_\be \Lambda_{\nu \beta}\sqrt{\frac{(\lambda_\nu+\omega)(J_\be b^2_\be -(\lambda_\be+\omega)(\mu+\omega))}{(\lambda_\be+\omega)(J_\nu b^2_\nu -(\lambda_\nu+\omega)(\mu+\omega))}},\nonumber
\eeq
defines the boost factor and $\epsilon(\om)$ is given in Eq.~\eqref{eq:inverse_eigen}.
Here $\Lambda_{\nu \be}$ are matrix elements of the $2\times2$ matrix
\beq
\Lambda = \left ( \begin{array}{cc} 0 & 1\\
1 &  0
\end{array} \right ).
\eeq

The equations (\ref{app:MFdensity}) and the relations between the different
spectral densities (\ref{app:densities}) for a given sublattice $\nu$  are formally the
same as in the homogeneous case $J_1=J_2$ (\ref{app:MFE}).



\begin{thebibliography}{99}
\bibitem{ions_foot}
The materials of the (RE)$_3$Pd$_{20}$X$_6$ type crystallize in cubic $Fm\bar{3}m$ structure, with
three RE ions per formula unit. Out of these, two occupy the 8c site whereas the third
occupies the 4a site, such that RE ions with two distinct environments exist.
%
Compared to the simpler two-ion model treated in the body of the paper, the Fermi-volume
considerations in the regime dominated by inter-moment exchange will be different -- due
to the odd number of $f$ electrons per unit cell they will be more akin to standard Kondo
lattices with a single moment per unit cell -- but the physics in the Kondo-dominated
regime will be qualitatively similar, as two distinct screening scales exist in
both cases.
\bibitem{coherence_foot}
There are various related definitions of a scale $\bTcoh$ in the literature. The one
proposed in Ref. \onlinecite{burdin1} is chosen such that $\bTcoh=T_K$ for a single Kondo
impurity.
\bibitem{entropy_foot}
In the slave-boson approach, the entropy of a free moment is $2\ln 2$
instead of $\ln 2$, due to the mean-field treatment of the Hilbert-space constraint.
\end{thebibliography}
\end{document}